\documentclass[preprint,authoryear,11pt]{elsarticle}

\usepackage{fancyhdr}
\usepackage{fullpage}
\usepackage{amsmath}
\usepackage{amsbsy}
\usepackage{amssymb}
\usepackage{amscd}
\usepackage{amsfonts}
\usepackage{supertabular}
\usepackage{graphics}
\usepackage{verbatim}
\usepackage{subfig}
\usepackage{epsfig}
\usepackage{xspace}
\usepackage{euscript}
\usepackage{alltt}
\usepackage{boxedminipage}
\usepackage{float}
\usepackage[colorlinks]{hyperref}
\usepackage{color}
\usepackage[all]{xy}
\usepackage{t1enc}
\usepackage{times,exscale}
\usepackage{graphicx,calc}
\usepackage[ruled,vlined]{algorithm2e}

\def\bR{{\mathbf{R}}}
\def\bx{{\mathbf{x}}}
\def\R{{\mathbb{R}}}

\newcommand{\norm}[1]{\| #1 \|}

\begin{document}

\begin{frontmatter}

\title{Coarse-graining Kohn-Sham Density Functional Theory}
\author[gatech]{Phanish Suryanarayana}
\author[caltech]{Kaushik Bhattacharya\corref{cor}}
\author[caltech]{Michael Ortiz}
\address[gatech]{College of Engineering, Georgia Institute of Technology, GA 30332, USA}
\address[caltech]{Division of Engineering and Applied Science, California Institute of Technology, CA 91125, USA}
\cortext[cor]{Corresponding Author (\it bhatta@caltech.edu) }

\begin{abstract}
We present a real-space formulation for coarse-graining Kohn-Sham Density Functional Theory that significantly speeds up the analysis of material defects without appreciable loss of accuracy. The approximation scheme consists of two steps. First, we develop a linear-scaling method that enables the direct evaluation of the electron density without the need to evaluate individual orbitals. We achieve this by performing Gauss quadrature over the spectrum of the linearized Hamiltonian operator appearing in each iteration of the self-consistent field method. Building on the linear-scaling method, we introduce a spatial approximation scheme resulting in a coarse-grained Density Functional Theory. The spatial approximation is adapted so as to furnish fine resolution where necessary and to coarsen elsewhere. This coarse-graining step enables the analysis of defects at a fraction of the original computational cost, without any significant loss of accuracy. Furthermore, we show that the coarse-grained solutions are convergent with respect to the spatial approximation. We illustrate the scope, versatility, efficiency and accuracy of the scheme by means of selected { examples}.
\end{abstract}


\end{frontmatter}

\section{Introduction} \label{Sec:Introduction}

Crystal defects play a critical role in determining macroscopic properties of solids even when present in small concentrations \citep{Phillipsbook}. For instance, vacancies, perhaps the simplest type of defect, are fundamental to phenomena like creep, spall and radiation ageing even when present in only parts per million.  Dislocations enable plasticity even though their densities are typically as small as 10$^{-8}$ per atomic row.   Stacking faults also influence plasticity and surface energies affect fracture. Further, small changes in composition often have a dramatic effect on mechanical properties because they segregate to defects. The reason that defects play such a profound role is that they bring together the chemical effects of the core, the discrete effects of the lattice and the long-range effects of the elastic fields.  Consider specifically a vacancy.  On the one hand, the nature of the core is determined by the chemistry of the dangling bonds.  On the other hand, the vacancy gives rise to displacements of atoms that decay at a polynomial rate, and this enables long-range interactions with other defects as well as macroscopicically applied fields. This coupling grows stronger in extended defects. Thus, a complete understanding of defects at physically relevant concentrations requires a simultaneous study of the details of the core as well as the long-range elastic fields.

This remains an outstanding challenge.  On the one hand, methods like Density Functional Theory (DFT) that are capable of accurately describing the { chemistry of the} defect core are much too complicated and expensive to use for defects with long-range fields.  On the other hand, methods capable of describing the long-range fields including atomistic and continuum methods, lack the fidelity and rely on empiricism to describe the core.   This challenge has prompted the development of multiscale approaches which seek to use DFT or tight-binding (TB) for the core, atomistic models for the near field and continuum for the far-field (e.g., \cite{Carter1999, Kaxiras2005, Kaxiras2006, Bernstein2009}).  Others have used parameter passing where detailed simulations are used to fit the parameters of a coarser model (e.g., \cite{Cuitino:2002p4605}). While these multiscale methods provide valuable insight, there is no seamless transition from DFT to TB or empirical potentials to continuum. Uncontrolled approximations resulting from these seams, the use of linear-response theory and kinematic assumptions like the Cauchy-Born hypothesis may render the methods unreliable. Furthermore, there is often no guarantee of convergence of the solutions provided by these methods to the full DFT solution.  

An alternative approach to multiscale modeling was initiated through the quasicontinuum method \citep{Tadmor1996}.  Here, one seeks to solve a single theory in the complete region, but using an adaptive numerical method where all the details are represented close to the core but only sampled farther away.  There is no additional modeling (uncontrolled approximations), and all the approximations are in the numerical scheme so that it converges to the detailed theory with increased resolution.  \cite{Tadmor1996} and subsequent refinements (see \cite{Miller:2009p4593} for a recent review) use an atomistic approach with empirical potentials as the only theory.  To that extent, it is unable to describe the full details of the defect core.

The Schr{\"o}dinger equation is fundamental for describing the quantum mechanical electronic structure of matter. However, the solution of the Schr{\"o}dinger equation is exceedingly expensive (scaling exponentially with number of atoms) and therefore impractical. A far-reaching reformulation of the problem was achieved by \cite{Hohenberg}, who proved the existence of a one-to-one correspondence between the ground-state electron density and the ground-state wavefunction of a many-particle system.  By this correspondence, the electron density replaces the many-body electronic wavefunction as the fundamental unknown field, thereby greatly reducing the dimensionality and computational complexity of the problem.  However, this theory requires an unknown but universal (independent of material) exchange and correlation functional.   Computationally convenient models have been developed, including the local density approximation (LDA) \citep{Kohn1965} and the generalized gradient approximation (GGA) \citep{Langreth,PerdewGGA}.

Traditional formulations of DFT follow \cite{Kohn1965} and solve for the orbitals \citep{PARSEC, VASP, Sterne, DFT++, CASTEP, ABINIT, Tsuchida, OCTOPUS, Phanish2010, Phanish2011}.  {  Since the number of orbitals is proportional to the number of atoms and they have to be mutually orthogonal, these formulations typically result in cubic-scaling with respect to the number of atoms}.  This places severe restrictions on the size of the systems which can be studied. To overcome this, a number of methods have been proposed that exhibit better scaling properties, with particular emphasis on linear-scaling \citep{Parrinello1992,Car,Goedecker, ONETEP,Barrault2007,Weinan2009,  Bowler2012}.    { See \cite{Goedecker} and \cite{Bowler2012} for exhaustive reviews of these methods.  Briefly, a vast majority of linear scaling methods avoid orbitals and instead reformulate DFT in terms of the so-called density matrix.  This is a diagonally dominant matrix and linear scaling is obtained by cutting off the decaying off-diagonal components.  Often a basis set made of localized Wannier functions is used to represent the density matrix, and the decay of off-diagonal components is equivalent to the decay of these functions.  The cutoff  is acceptable in insulators where the band gap results in a strong exponential decay.  In metals however, one only has polynomial decay in general.  While, it has been shown that exponential decay may exist in idealized examples at finite temperatures \citep{Goedecker}, a mathematical understanding of the decay properties and consequently the accuracy of these methods remain unclear.  Therefore,} the development of a linear-scaling method for metallic systems remains an open problem \citep{Cances2008}.

Even with a linear scaling DFT method, the study of crystal defects at realistic concentrations remains a daunting problem. 
\cite{Gavini:2007p340} extended the quasicontinuum approach to electronic structure by implementing it on a simplified version of DFT known as orbital-free DFT (OFDFT).  OFDFT models the kinetic energy of the electrons instead of computing it explicitly using wave functions, and is limited in its applicability to free electron metals like Aluminum.  \cite{Gavini:2007p340} use a nested discretization to fully resolve both the atomic displacements and electronic density close to the core, but sample it elsewhere. They demonstrated the efficacy of their method by exhibiting computational savings of a factor of $10^3$ or more, and also showed through examples how the physics can change depending on the size of the computational cell (concentration of defects).  Unfortunately, it is not possible to directly extend their idea to the standard formulation of DFT for three main reasons.  First, orbitals are subject to orthogonality, which is a global constraint and difficult to coarse-grain.  From a practical point of view, orthogonality means that orbitals oscillate extremely rapidly and this makes  the quasicontinuum representation difficult.  Second, they are not periodic even in homogenous systems (instead, they are Bloch-Floquet waves).  Finally, one has as many orbitals as electrons, and thus the method would scale poorly even with a quasicontinuum representation of each orbital.

In this paper, we present a method that overcomes these difficulties.  Specifically, we present a formulation (which we call CGDFT) that seamlessly coarse-grains DFT by recourse to controlled numerical approximations without the introduction of new or spurious physics.  We accomplish this through a series of steps.  First, we reformulate DFT in such a manner that eliminates the need to explicitly compute orbitals.  Instead of computing the orbitals, and then using them to compute the quantities of interest including Fermi energy, electron density and band structure energy, we use integral representations of these quantities in terms of the projected density of states.  Thus, the theory is formulated in terms of smoothly varying quantities that are amenable to coarse-graining.  Second, we perform numerical Gauss quadrature on the spectrum of the Hamiltonian and use operator theory to evaluate these integrals.   Third, we develop an algorithm that truly scales linearly with the number of atoms.  Together we refer to these steps as the Linear Scaling Spectral Gauss Quadrature (LSSGQ) method.  As a final step, the spatial approximation is adapted so as to furnish fine resolution where necessary and to coarsen elsewhere. This coarse-graining step enables the analysis of defects at a fraction of the original computational cost, without any significant loss of accuracy. Furthermore, we show that the CGDFT solutions are convergent with respect to the spatial approximation. These properties render CGDFT to be highly transferable, efficient and accurate.

{The fundamental difference of the LSSGQ formulation from existing linear-scaling methods is that no explicit cutoff is employed.  The formulation resembles the Fermi operator expansion (FOE) method \citep{Goedecker,Bowler2012}, but there are some notable differences. Importantly, instead of expanding the density matrix in terms of a polynomial basis or rational functions, we use Gauss quadratures to evaluate diagonal elements of the density matrix and other quantities of interest. For this reason, LSSGQ does not involve the representation of the complete density matrix.  Instead, only the components of interest are extracted.  Further, the Fermi energy is not needed as an input to the method. This enables significant savings since the method does not have to be iterated over for the calculation of the Fermi energy. Additionally, symmetries in the system can be easily utilized to significantly reduce the computational effort.  Finally the proposed method offers adaptivity through controllable accuracy in space, a useful feature for multiscale and coarse-graining formulations. The LSSGQ fomulation also has similaries to the recursion method used in tight-binding \citep{HAYDOCK:1972p4639, HAYDOCK:1975p4638,Haydock1980}. In the recursion method, a continued fraction representation of the projected density of states is developed and quantities of interest are obtained by integration over it. However, we proceed differently and directly calculate the Gauss quadrature rules, which is a much more accurate and stable process.}

The remainder of this paper is organized as follows. In Section \ref{Sec:KS-DFT}, we provide a brief introduction to DFT and in particular to the Kohn-Sham method. { This includes a heuristic overview of our reformulation in Section \ref{Sec:Heuristics} }. Next, we provide some mathematical background to spectral theory and Gauss quadrature in Section \ref{Sec:MathematicalBackground}. Subsequently, we present and validate the LSSGQ formulation in Section \ref{Sec:LSSGQ}. In Section \ref{Sec:CGDFT}, we describe CGDFT, and validate it through select examples. Finally, we conclude in Section \ref{Sec:Conclusions}.

%%%%%%%%%%%%%%%%%%%%%%%%%%%%%%%%%%%%%%%%%%%%%%%%%%%%%%%%%%%%%%%%%%%%%%%%%%%%%%%%%%%%

\section{Kohn-Sham Density Functional Theory} \label{Sec:KS-DFT}

{ 
\subsection{Orbital formulation} \label{Sec:Tradtional}
Consider a system of $M$ atoms with $N_e$ electrons. DFT is a theory which can be used to find the ground-state energy of the system, the ground-state electron density and the equilibrium position of the nuclei.  In our presentation, we ignore electron spin and assume that $N_e$ is even for clarity, but note that all the methods developed in the paper may easily be extended to include spin. Let $\bR = \{\bR_1, \bR_2, \ldots, \bR_M \}$ denote the positions of the nuclei with charges $\{Z_1, Z_2, \ldots, Z_M \}$ respectively.  We follow \cite{Kohn1965} and introduce orbitals $\Psi =$ $\{\psi_{1}$, $\psi_{2}$, $\ldots$, $\psi_{N_{e}/2}$ \} for the electrons.  Then, the energy of the system is written as 
\begin{equation} \label{Eqn:Energy:orbitals}
\mathcal{E} (\Psi, \bR) = -\sum_{n=1}^{N_e/2} \int_{\R^3} \psi_{n}^{*}(\bx) \nabla^2 \psi_{n}(\bx) \, \mathrm{d\bx} + E_{\rm xc}(\rho) + E_{\rm H}(\rho) + E_{\rm ext}(\rho,\bR) + E_{\rm zz}(\bR) \,,
\end{equation}
where the electron density
\begin{equation} \label{Eqn:ElectronDensity}
\rho(\bx) = 2 \sum_{n=1}^{N_e/2} |\psi_{n}(\bx)|^2.
\end{equation}
The first term in Eqn. \ref{Eqn:Energy:orbitals} represents the kinetic energy of the non-interacting electrons.   The second term, $E_{\rm xc}(\rho)$, denotes the exchange-correlation energy. For definiteness, we adopt the local density approximation (LDA)\citep{Kohn1965}
\begin{equation}
E_{xc} = \int_{{\mathbb R}^3} \rho \epsilon_{xc} (\rho) \, \mathrm{d\bx}
\end{equation}
where $\epsilon_{xc} (\rho)$ is the exchange-correlation energy density. The final three terms are electrostatic in nature,
\begin{eqnarray}
E_{\rm H}(\rho) & = & \frac{1}{2} \int_{\R^3} \int_{\R^3} \frac{\rho(\bx)\rho(\bx')}{|\bx - \bx'|} \,\mathrm{d\bx} \, \mathrm{d\bx'}\,, \label{eq:EH} \\
E_{\rm ext} (\rho,\bR) & = & \int_{\R^3} \rho(\bx) \left( \sum_{I=1}^{M} \frac{Z_{I} }{|\bx - \bR_{I}|}\right) \, \mathrm{d\bx} \,, \\
E_{\rm zz}(\bR) & = & \frac{1}{2} \sum_{I=1}^{M} \sum_{\begin{subarray}{l} J=1 \\J \neq I \end{subarray}}^{M} \frac{Z_{I} Z_{J}}{|\bR_{I}-\bR_{J}|} \, .\label{eq:EZZ}
\end{eqnarray}
$E_{\rm H}$ is known as the Hartree energy and is the classical electrostatic interaction energy of the electron density, $E_{\rm ext}$ is the electrostatic interaction energy between the electron density and nuclear charges, and $E_{\rm zz}$ is the repulsive energy between the nuclei. 

For a given position of the nuclei $\bR$, the electronic ground-state energy of the system is obtained as the solution to the variational principle. 
\begin{equation} \label{varprinc}
\mathcal{E}_0 (\bR) = \inf_{\Psi  }\mathcal{E}(\Psi,\bR)
\end{equation}
subject to the orthonormality constraints
\begin{equation}\label{orthonormal_constraints}
\int_{\R^3} \psi_{i}^{*}(\bx) \psi_{j}(\bx) \,\mathrm{d\bx} = \delta_{ij} \,, \qquad i,j =  1, 2, \ldots, N_{e}/2 \, .
\end{equation}  }

It is convenient to rewrite the electrostatic terms by introducing the electrostatic potential $\phi$ as the solution to the Poisson equation (cf e.g. \cite{DFT++,Phanish2010})
\begin{equation} \label{Eqn:Poisson:ElectrostaticPotential}
-\frac{1}{4\pi} \nabla^{2}\phi(\bx,\bR) = \rho(\bx) + b(\bx,\bR) 
\end{equation}
where $b(\bx,\bR) = \sum_{J=1}^{M} b_{J}(\bx,\bR_J)$ denotes the total charge density of the nuclei, with $b_{J}(\bx,\bR_J)$ representing the regularized charge density of the $J^{th}$ nucleus. { Thereafter, the electrostatic energies may be rewritten as  
\begin{eqnarray}
E_{\rm H} + E_{\rm ext} + E_{\rm zz} &=& - \frac{1}{8 \pi} \int_{\R^3} |\nabla\phi(\bx,\bR)|^2\mathrm{d\bx} 
+\int_{\R^3} (\rho(\bx)+b(\bx,\bR))\phi(\bx,\bR)\, \mathrm{d\bx} \\
&&  \quad - \frac{1}{2}\sum_{J=1}^{M} \int_{\R^3} \int_{\R^3} \frac{b_J(\bx,\bR_J) b_J(\bx',\bR_J)}{|\bx-\bx'|} \, \mathrm{d\bx d\bx'} \nonumber \, .
\end{eqnarray} 
where the last term denotes the self energy of the nuclei.}

The Euler-Lagrange equations of the constrained variational principle (Eqns. \ref{varprinc} and \ref{orthonormal_constraints}) gives rise to the nonlinear eigenvalue problem
\begin{equation} \label{Eqn:EigValue:Zero}
\mathcal{H} \psi_{n} = \lambda_{n} \psi_{n} \,, \quad n = 1, 2, \ldots N_e/2 
\end{equation}
where the Hamiltonian
\begin{equation} \label{Eqn:Hamiltonian}
\mathcal{H} = -\frac{1}{2}\nabla^2 + V_{xc}(\rho) + \phi(\bx,\bR)
\end{equation}
is a self-adjoint operator with eigenvalues $\lambda_{n}$ (ordered so that $\lambda_1 \le \lambda_2 \le \dots)$ and $V_{xc}(\rho) = \frac{\delta E_{xc}(\rho)}{\delta \rho (\bx)}$.  { The problem is nonlinear since the Hamiltonian depends on $\rho$ (through $V_{xc}(\rho)$ and $\phi(\bx,\bR)$) which in turn depends on $\psi_n$. Therefore, this problem is typically solved by a fixed point iteration with respect to the electron density, known as the self-consistent field (SCF) method (cf. e.g. \cite{Martin2004}). In each iteration of the SCF method, the electron density is calculated by solving for the eigenfunctions $\psi_n$ corresponding to the lowest $N_e/2$ eigenvalues $\lambda_n$, and then using Eqn. \ref{Eqn:ElectronDensity}. This is indeed equivalent to the variational problem (cf. e.g. \cite{Phanish2010}). 

The electronic ground-state energy is obtained by substituting the solution of Eqn. \ref{Eqn:EigValue:Zero} in Eqn. \ref{Eqn:Energy:orbitals}.   We can then use the eigenvalue problem given by Eqn. \ref{Eqn:EigValue:Zero} (left multiply by $\psi_n^{*}$ and integrate) to show that this ground-state energy is}
\begin{eqnarray} \label{groundstate}
\mathcal{E}_0 (\bR) & = & 2 \sum_{n=1}^{N_e/2} \lambda_{n}  + E_{xc}(\rho) + \frac{1}{2} \int_{\mathbb{R}^3} (b(\bx,\bR)-\rho(\bx)) \phi(\bx,\bR) \, \rm{d\bx}  \nonumber \\
& & - \int_{\mathbb{R}^3} V_{xc}(\rho) \rho(\bx) \, \rm{d\bx} - \frac{1}{2}\sum_{J=1}^{M} \int_{\R^3} \int_{\R^3} \frac{b_J(\bx,\bR_J) b_J(\bx',\bR_J)}{|\bx-\bx'|} \, d\bx d\bx'   \label{Eqn:KS:Energy}
\end{eqnarray}
where $U_{band} = 2 \sum_{n=1}^{N_e/2} \lambda_{n}$ denotes the band structure energy. 

{  In this orbital approach, we solve the nonlinear eigenvalue problem (Eqn. \ref{Eqn:EigValue:Zero}) for the first $N_e/2$ eigenvalues and the corresponding eigenfunctions. Subsequently, the electronic ground-state energy can be evaluated using Eqn. \ref{groundstate}. This formulation of DFT requires an effort scaling as $\mathcal{O}(N_e^3)$ (\cite{Goedecker}), which restricts the size of systems that can be studied. Further, the orbitals are not amenable to coarse-graining since they are extremely oscillatory and are not periodic even in periodic systems. Finally, a quasicontinuum representation of each orbital would scale poorly since the number of orbitals increases with the number of electrons.

\subsection{Heuristics of the proposed reformulation} \label{Sec:Heuristics}
The key idea behind our reformulation of DFT is to note that the ground-state energy (Eqn. \ref{groundstate}) does {\it not} depend on individual eigenvalues and eigenfunctions, but only on certain sums:
\begin{equation}
U_{band} = 2 \sum_{n=1}^{N_e/2} \lambda_{n}, \quad \rho (\bx) = 2 \sum_{n=1}^{N_e/2} |\psi_{n} (\bx)|^2.
\end{equation}
So it seems natural to rewrite these sums by sampling and weighting:
\begin{equation}
U_{band} = 2 \sum_{k=1}^{K} w_k \bar{\lambda}_k \,, \quad \rho(\bx) = 2 \sum_{k=1}^{K} w_k |\bar{\psi}_k(\bx)|^2 
\end{equation} 
for appropriately chosen sampling `eigenvalues' $\bar{\lambda}_k$, `eigenfunctions' $\bar{\psi}_k$ and weights $w_k$.  We seek to find a way of choosing the sampling points optimally and computing the weights efficiently. Particularly, we would like $K$ independent of $N_e$ and $K \ll N_e/2$ (for large $N_e$) such that the overall effort in evaluating these sums is ${\mathcal O}(N_e)$. Since the number of electrons is proportional to the number of atoms, the computational effort will scale as  ${\mathcal O}(M)$.  

We accomplish this in four steps. We outline these here, and develop them precisely in Sections \ref{Sec:LSSGQ} after recalling some necessary mathematical preliminaries in Section \ref{Sec:MathematicalBackground}.  First, we introduce the orbital occupation function
\begin{equation} \label{Eqn:StepFunction}
g(\lambda,\lambda_{f}) =
\begin{cases}
1 , & \text{if  } \lambda \leq \lambda_{f} \\
0 , & \text{otherwise  }
\end{cases}
\end{equation}
and rewrite $U_{band}$ and $\rho(\bx)$ as
\begin{equation} \label{Eqn:Ubandrho}
U_{band} = 2 \sum_{n}^{} \lambda_n g(\lambda_n, \lambda_f), \quad 
\rho(\bx) = 2 \sum_{n}^{} g(\lambda_n, \lambda_f) |\psi_{n}(\bx)|^2.
\end{equation}
Above  $\lambda_{f} = \lambda_{N_e/2}$ is called the Fermi energy.  This is unknown \emph{a priori} but can be solved through the constraint
\begin{equation} \label{Eqn:FermiEnergyConstraint}
N_e = 2 \sum_n g(\lambda_{n},\lambda_{f}) .
\end{equation}
Second, we rewrite the sums in Eqns. \ref{Eqn:Ubandrho} and \ref{Eqn:FermiEnergyConstraint} as spectral integrals of the form
\begin{equation}
I[f] = \int_{\sigma({\mathcal H})} f(\lambda) \mathrm{d\sigma}(\lambda)
\end{equation}
where $\sigma({\mathcal H})$ is the spectrum of the Hamiltonian ${\mathcal H}$ and d$\sigma(\lambda)$ is the appropriate spectral measure.  We provide an introduction to spectral theory in Section \ref{Sec:LSSGQ:SpectralTheory} and develop integral representations in Section \ref{Sec:LSSGQ:IntegralRepresentations}.  Third, we evaluate these spectral integrals using Gauss quadratures
\begin{equation}
I[f] \approx \sum_{k=1}^{K} w_k f(\lambda_k).
\end{equation}
Finally, we evaluate the quadrature nodes $\{ \lambda_k \}_{k=1}^{K}$ and weights $\{ w_k \}_{k=1}^{K}$ using an efficient Lanczos type iteration.
We provide an introduction to Gauss quadratures and the Lanczos type iteration in Section \ref{Sec:LSSGQ:GaussQuadrature} and a precise algorithm in Section \ref{Sec:LSSGQ:Formulation}. We discuss the scaling and performance of the method in Section \ref{Sec:LSSGQ:ScalingPerformance}. 

The reformulation as described above overcomes the difficulties encountered when attempting to coarse-grain the orbital formulation. Notably, the spectral quadrature points and  weights can be formulated as local real space variables (at each spatial discretization point), and these are smoothly varying quantities.  This enables us to coarse-grain the problem in Section \ref{Sec:CGDFT}. }

\subsection{Atomic positions}
To calculate the ground-state energy of the system, we need to further minimize the energy with respect to the positions of the nuclei. To do so, we take the first variation of $\mathcal{E}_0(\bR)$ with respect to $\bR$ to obtain the forces on the nuclei 
\begin{equation} \label{Eqn:Force:Nuclei}
f_J = - \int_{\R^3} \frac{\partial b_J(\bx,\bR_J)}{\partial \bR_J}\left(\phi(\bx,\bR)-\phi_J(\bx,\bR_J)\right) \, \rm{d\bx}
\end{equation}
where $\phi_J(\bx,\bR_J) = \int_{\R^3} \frac{b_J(\bx',\bR_J)}{|\bx-\bx'|} \, \rm{d\bx'}$ and $f_J$ represents the force on the $J^{th}$ nucleus. In the special case where we have $b_J(\bx,\bR_J)=-Z_J \delta(\bx-\bR_J)$, we obtain
\begin{equation} \label{Eqn:Force:Nuclei_AllElectron}
f_J = Z_J \nabla (\phi(\bx,\bR)-\phi_J(\bx,\bR_J)) \bigg|_{\bx=\bR_J} .
\end{equation}
This result is commonly referred to as the Hellmann-Feynman theorem (cf. e.g. \cite{Finnis2003}).

{ 
\subsection{Pseudopotential Approximation}
The core states are localized in the vicinity of the nucleus leading to oscillations of the valence orbitals in this region. Irrespective of the basis set used, a large number of basis functions are required to capture these oscillations. Additionally, the tightly bound core electrons are chemically inactive and hence have a negligible contribution towards determining physical properties. In view of this, the core electrons are eliminated and an effective nuclear potential is introduced to describe the effect of the core electrons, resulting in nodeless pseudo-orbitals. This amounts to replacing the all-electron potential $V_{\rm ext}(\bx,\bR) = \sum_{I=1}^{M} \frac{Z_{I} }{|\bx - \bR_{I}|} $ with an effective potential $V_{\rm ext}^{\rm PS}(\bx,\bR)$. 

The pseudopotentials can be broadly classified as either local or non-local based on their spatial dependence. The local pseudopotentials are explicit functions $V_{\rm ext}^{\rm PS}(\bx,\bR)$. They can be incoporated into our formulation by letting $b(\bx,\bR)=\frac{-1}{4\pi} \nabla^2 V_{\rm ext}^{PS}$. In contrast, non-local pseudopotentials are operators on the orbital with angular momentum dependence and are designed to accurately reproduce the scattering properties of the all-electron potential. These include norm conserving \citep{Bachelet,Rappe,Troullier} and ultrasoft pseudopotentials \citep{Vanderbilt}, which are usually employed in the Kleinman-Bylander form \citep{Kleinman}. For the examples is this paper, we utilize a local pseudopotential termed as the `Evanescent Core' pseudopotential \citep{Fiolhais1995}. However, it should be noted that the methods proposed in this work are applicable even with the choice of non-local pseduopotentials.  }

{  \subsection{Finite temperature extension} }
In previous discussion of the Kohn-Sham method, we have tacitly assumed a temperature of absolute zero. The method can be easily extended to finite temperatures (cf. e.g. \cite{Parr1989}). The Kohn-Sham problem for the electronic ground-state takes the form
\begin{eqnarray}
\mathcal{H} \psi_{n} & = & \lambda_{n} \psi_{n}\,, \quad \mathcal{H} = -\frac{1}{2}\nabla^2 + V_{xc}(\rho) + \phi(\bx,\bR) \,, \nonumber \\
-\frac{1}{4\pi} \nabla^{2}\phi(\bx,\bR) & = & \rho(\bx) + b(\bx,\bR) \,,  \nonumber \\
\rho(\bx) & = & 2 \sum_n g(\lambda_{n},\lambda_{f}) |\psi_{n}(\bx)|^2 \,,  \label{Eqn:KS:Problem:EulerLagrange} \\
N_e & = & 2 \sum_n g(\lambda_{n},\lambda_{f})  \nonumber 
\end{eqnarray}
where the orbital occupation is now allowed to take fractional values as defined by the Fermi-Dirac distribution
\begin{equation}
g(\lambda,\lambda_f) = \frac{1}{1+\exp(\frac{\lambda-\lambda_f}{\sigma})} .
\end{equation}
Above, $\sigma = k_B \theta$, $k_B$ being the Boltzmann constant and $\theta$ is the absolute temperature. The ground-state  Helmholtz free energy
\begin{equation} \label{Eqn:FreeEnergy}
\mathcal{F} = \mathcal{E}_{\sigma} - \theta \mathcal{S}
\end{equation}
where 
\begin{eqnarray}
\mathcal{E}_{\sigma} & = & 2 \sum_{n} g(\lambda_{n},\lambda_{f}) \lambda_{n}  + E_{xc}(\rho) + \frac{1}{2} \int_{\mathbb{R}^3} (b(\bx,\bR)-\rho(\bx)) \phi(\bx,\bR) \, \rm{d\bx}  \nonumber \\
& & - \int_{\mathbb{R}^3} V_{xc}(\rho) \rho(\bx) \, \rm{d\bx} - \frac{1}{2}\sum_{J=1}^{M} \int_{\R^3} \int_{\R^3} \frac{b_J(\bx,\bR_J) b_J(\bx',\bR_J)}{|\bx-\bx'|} \, d\bx d\bx' \nonumber 
\end{eqnarray} 
is the finite temperature counterpart of the ground-state energy at absolute zero ($\mathcal{E}_{0}$) and 
\begin{equation}
\mathcal{S} = - 2 k_{B} \sum_{n} \left[ g(\lambda_n,\lambda_f) \log g(\lambda_n,\lambda_f) + (1-g(\lambda_n,\lambda_f)) \log (1-g(\lambda_n,\lambda_f)) \right] 
\end{equation}
is the entropy resulting from the fractional orbital occupations. The ground-state position of the nuclei can be found be equilibriating Eqn. \ref{Eqn:Force:Nuclei} \citep{Weinert1992}. Using the finite temperature calculation, it is also possible to to get an accurate extrapolation for the ground-state energy at absolute zero \citep{Gillan1989}
\begin{equation} \label{Eqn:FiniteTemperatureExtrapolation}
\mathcal{E}_0 \approx \frac{1}{2} \left( \mathcal{E}_{\sigma} + \mathcal{F} \right).
\end{equation}

%%%%%%%%%%%%%%%%%%%%%%%%%%%%%%%%%%%%%%%%%%%%%%%%%%%%%%%%%%%%%%%%%%%%%%%%%%%%%%%%%%%%%%%%%%%%%%%%%%%%%%%%%%%%%%%%%%%%5

\section{Mathematical Background} \label{Sec:MathematicalBackground}

In this section, we recall a few mathematical tools which we exploit to develop our methods.  We refer the reader to \cite{Rudin1991} for further details on spectral theory and \cite{Golub2010} for Gauss quadratures.

\subsection{Spectral theory} \label{Sec:LSSGQ:SpectralTheory}

Let $\mathcal{H}$ be a self-adjoint linear operator on a finite-dimensional Hilbert space $H$ with inner product $(.,.)$ and norm $\norm{.}$.  Then, according to the spectral theorem, we have a unique resolution of the identity $\mathcal{E}$ that satisfies
\begin{equation} \label{Eqn:SD:T}
\mathcal{H} = \int_{\sigma(\mathcal{H})} \lambda \, \mathrm{d\mathcal{E} (\lambda)}
\end{equation} 
where $\sigma(\mathcal{H}) = \{\lambda_1, \lambda_2, \ldots , \lambda_{N_d} \}  \subset \R $ is the spectrum of $\mathcal H$.  Consequently, for any bounded function $f$ on $\sigma(\mathcal{H})$,
\begin{equation}\label{Eqn:SD:fT}
f(\mathcal{H}) = \int_{\sigma(\mathcal{H})} f(\lambda) \, \mathrm{d\mathcal{E} (\lambda)}.
\end{equation}
{Note that  Eqn. \ref{Eqn:SD:fT} can also be written using Dirac's bra-ket notation as $f(\mathcal{H}) = \sum_{n=1}^{N_d} f(\lambda_n) |\psi_n \rangle \langle \psi_n|$, where $\psi_n$ denotes the eigenfunction corresponding to the eigenvalue $\lambda_n$.}  We assume that the eigenvalues are ordered such that  $\lambda_1 \leq \lambda_2 \leq \ldots \leq \lambda_{N_d}$. 

Let $\{\eta_p\}_{p=1}^{N_d}$ be an orthonormal basis of $H$ so that for any $\zeta \in H$, $\zeta = \sum_{p=1}^{N_d} \zeta_p \eta_p$. The eigenfunctions can therefore be expressed as $\psi_n = \sum_{p=1}^{N_d} \psi_{n,p} \eta_p $.  Thereafter, we have the following representation of the measure $\mathcal{E}_{\zeta,\zeta}(\lambda)$ on $\sigma(\mathcal{H})$
\begin{equation}
\mathcal{E}_{\zeta,\zeta}(\lambda) = (\mathcal{E}(\lambda)\zeta,\zeta) =
\begin{cases}
0, & \text{if  } \lambda < \lambda_1 \\
\sum_{n=1}^{m} \sum_{p=1}^{N_d} \sum_{q=1}^{N_d} \psi_{n,p} \psi_{n,q} \zeta_{p} \zeta_{q}, & \text{if  } \lambda_{m}\leq\lambda<\lambda_{m+1} \\
\sum_{n=1}^{N_d} \sum_{p=1}^{N_d} \sum_{q=1}^{N_d} \psi_{n,p} \psi_{n,q} \zeta_{p} \zeta_{q}, & \text{if  } \lambda_{N_d}<\lambda
\end{cases} \, .
\end{equation}
In the special case of $\zeta=\eta_k$, it reduces to
\begin{equation}
\mathcal{E}_{\eta_k,\eta_k}(\lambda) =
\begin{cases}
0, & \text{if  } \lambda < \lambda_1 \\
\sum_{n=1}^{m} |\psi_{n,k}|^{2}, & \text{if  } \lambda_{m}\leq\lambda<\lambda_{m+1} \\
\sum_{n=1}^{N_d} |\psi_{n,k}|^{2}, & \text{if  } \lambda_{N_d}<\lambda .
\end{cases}
\end{equation}
Therefore for any $\zeta \in H$,
\begin{equation}\label{Eqn:SD:fT:measure}
(f(\mathcal{H}) \zeta, \zeta) = \int_{\sigma(\mathcal{H})} f(\lambda) \, \mathrm{d\mathcal{E}_{\zeta,\zeta}(\lambda)} = \int_{a}^{b} f(\lambda) \, \mathrm{d\mathcal{E}_{\zeta,\zeta}(\lambda)} \,,
\end{equation}
for $a=\lambda_1$ and $b=\lambda_{N_d}$. {It is worth noting that $\mathrm{d\mathcal{E}_{\zeta,\zeta}(\lambda)}$ represents the projected density of states of $\mathcal{H}$.}

\subsection{Gauss quadrature} \label{Sec:LSSGQ:GaussQuadrature}

Let $\mathcal{P}$ be the space of real polynomials and $\mathcal{P}_{K}$ be a subspace of $\mathcal{P}$ consisting of polynomials of degree $K$. We define an inner product (relative to the measure $\mathcal{E}_{\zeta,\zeta}$) of two polynomials $p, q \in \mathcal{P}$ as
\begin{equation}
\langle p,q \rangle_{\zeta} = \int_{a}^{b} p(\lambda) q(\lambda) \, \mathrm{d\mathcal{E}_{\zeta,\zeta} (\lambda)}
\end{equation}
with norm
\begin{equation}
\norm{p}_{\zeta} = \left( \int_{a}^{b} p^{2}(\lambda) \, \mathrm{d\mathcal{E}_{\zeta,\zeta} (\lambda)} \right)^{\frac{1}{2}}.
\end{equation}
Here, we are interested in approximating integrals of the form ({Eqn. \ref{Eqn:SD:fT:measure}})
\begin{equation} \label{Eqn:SampleIntegral}
I[f] = \int_{a}^{b} f(\lambda) \, \mathrm{d\mathcal{E}_{\zeta,\zeta}(\lambda)}
\end{equation}
using a quadrature rule. To do so, we approximate the function $f(\lambda)$ with an interpolation polynomial
\begin{equation}\label{Eqn:Interpolation:f}
f(\lambda) \approx \sum_{k=1}^{K} f(\lambda_k^{\zeta}) l_k^{\zeta}(\lambda)
\end{equation}
where $\{ \lambda_k^{\zeta} \}_{k=1}^{K}$ are the nodes/interpolation points and $l_k^{\zeta}(\lambda)$ is the Lagrange polynomial
\begin{equation}
l_k^{\zeta}(\lambda) = \prod_{j=1,j \neq k} \frac{\lambda-\lambda_j^{\zeta}}{\lambda_k^{\zeta}-\lambda_j^{\zeta}}.
\end{equation}
Thereafter, we obtain the quadrature formula
\begin{equation}\label{Eqn:QuadratureFormula}
I[f] \approx \sum_{k=1}^{K} w_k^{\zeta} f(\lambda_k^{\zeta})
\end{equation}
where
\begin{equation}
w_k^{\zeta} = \int_{a}^{b} l_k^{\zeta}(\lambda) \mathrm{d \mathcal{E}_{\zeta,\zeta}(\lambda)}
\end{equation}
are the weights of the quadrature rule. In order to obtain the quadrature rule of highest degree, we further consider the nodes $\{ \lambda_k \}_{k=1}^{K}$ as unknowns. This results in a quadrature rule of degree $2K-1$ i.e. any $p \in \mathcal{P}_{2K-1}$ is integrated exactly, commonly referred to as Gauss quadrature.

To efficiently evaluate the nodes and weights of the Gauss quadrature rule, we generate a sequence of orthonormal polynomials (with respect to the measure $\mathcal{E}_{\zeta,\zeta}$) $\{ \hat{p}_k \}_{k=0}^{K}$ through the three-term recurrence relationship
\begin{eqnarray}
b_{k+1} \hat{p}_{k+1}(\lambda) = (\lambda-a_{k+1}) \hat{p}_k(\lambda) - b_k \hat{p}_{k-1}(\lambda) \,, \quad k=0, 1, \ldots, K-1 \nonumber \\
\hat{p}_{-1}(\lambda) = 0 \,, \quad \hat{p}_{0}(\lambda) = 1 \,, \quad b_0 = 1 \label{Eqn:RecurrenceRelation}
\end{eqnarray}
where
\begin{equation}
a_{k+1} = \langle \lambda \hat{p}_k, \hat{p}_k \rangle_{\zeta} \,, \quad k=0, 1, \ldots, K-1
\end{equation}
and $\hat{b}_k$ is computed such that $\norm{\hat{p}_k}_{\zeta} = 1 \,, \, k=0, 1, \ldots, K$. Corresponding to these orthonormal polynomials, there is a tridiagonal Jacobi matrix $\hat{J}_K$ of dimension $K$
\begin{eqnarray} \label{Eqn:MatrixJK}
\hat{J}_K =  \left( \begin{array}{ccccc}
a_1 & b_1 & & & \\
b_1 & a_2 & b_2 & & \\
 & \ddots & \ddots & \ddots & \\
 & & b_{K-2} & a_{K-1} & b_{K-1} \\
 & & & b_{K-1} & a_K
\end{array} \right)
\end{eqnarray}
Let us denote $\hat{P}_K(\lambda) = (\hat{p}_0(\lambda), \hat{p}_1(\lambda), \ldots, \hat{p}_{K-1}(\lambda))^{T}$. Then the three term recurrence relation given by Eqn.~\ref{Eqn:RecurrenceRelation} can be written compactly as
\begin{equation}
\lambda \hat{P}_K(\lambda) = \hat{J}_K \hat{P}_K(\lambda) + b_K \hat{p}_K(\lambda) e_{K}
\end{equation}
where $e_{K}$ is the last column of the identity matrix of dimension $K$. It follows that the eigenvalues of $\hat{J}_K$ (which are also the zeros of $\hat{p}_K(\lambda)$) are the nodes $\{\lambda_k\}_{k=1}^{K}$ of the Gauss quadrature rule and the weights $\{w_k\}_{k=1}^{K}$ are the squares of the first elements of the normalized eigenvectors.

\section{Linear-Scaling Spectral Gauss Quadrature Method} \label{Sec:LSSGQ}

In this section, we describe the formulation, implementation and validation of the proposed linear-scaling spectral Gauss quadrature (LSSGQ) method. The outline of this section is as follows. We first develop integral representations for the quantities of interest in Section \ref{Sec:LSSGQ:IntegralRepresentations}. Subsequently, we describe the LSSGQ method in Section \ref{Sec:LSSGQ:Formulation} and discuss its scaling and performance in Section \ref{Sec:LSSGQ:ScalingPerformance}. Next, we discuss the LSSGQ method in the context of the finite-difference approximation in Section \ref{Sec:LSSGQ:FD}. Finally, we validate the LSSGQ method through examples in Section \ref{Sec:LSSGQ:NumericalExamplesValidation}. The connections of the LSSGQ method with the recursion method (\cite{Haydock1980}) and the Pad{\'e} approximation are discussed in Appendix \ref{Sec:RM}.

\subsection{Integral representations} \label{Sec:LSSGQ:IntegralRepresentations}

Consider a spatial discretization of our domain with $N_d$ basis functions.  Let $\mathcal{H}$ be the (linearized) Hamiltonian (Eqn. \ref{Eqn:Hamiltonian}) with a fixed $\rho$ and $\phi$.   In order to solve the Kohn-Sham problem, we need to evaluate the Fermi energy, updated electron density, band structure energy and entropy.  We begin by providing an integral representation of these quantities over the spectrum of $\mathcal{H}$.

\subsubsection{Fermi energy} \label{SubSubSec:FermiEnergy}
The Fermi energy is calculated by solving for the constraint
\begin{equation} \label{Eqn:FermiEnergyConstraintLDA}
N_e = 2 \sum_{n=1}^{N_d} g(\lambda_{n},\lambda_{f}).
\end{equation}
 Since $\norm{\psi_{n}} = \sum_{p=1}^{N_d}|\psi_{n,p}|^{2} = 1$, it follows that
\begin{eqnarray}
\sum_{n=1}^{N_d} g(\lambda_{n},\lambda_{f}) & = & \sum_{p=1}^{N_d} \sum_{n=1}^{N_d} g(\lambda_{n},\lambda_{f}) |\psi_{n,p}|^2 \nonumber \\
& = & \sum_{p=1}^{N_d} \int_{a}^{b} g(\lambda,\lambda_{f}) \mathrm{d \mathcal{E}_{\eta_p,\eta_p}(\lambda)} \label{Eqn:Equiv:Fermi}.
\end{eqnarray}
Therefore, Eqn.~\ref{Eqn:FermiEnergyConstraintLDA} can be rewritten as
\begin{equation}
N_{e} = 2 \sum_{p=1}^{N_d} \int_{a}^{b} g(\lambda,\lambda_{f}) \mathrm{d \mathcal{E}_{\eta_p,\eta_p}(\lambda)}.
\end{equation}

\subsubsection{Electron density} \label{SubSubSec:ElectronDensity}
The electron density at any point $\bx_0$ is
\begin{eqnarray}
\rho(\bx_0) & = & 2 \sum_{n=1}^{N_d} g(\lambda_n,\lambda_f) |\psi_{n}(\bx_0)|^{2} \nonumber \\
& = & 2 \sum_{n=1}^{N_d} \sum_{p=1}^{N_d} \sum_{q=1}^{N_d} g(\lambda_n,\lambda_f) \psi_{n,p} \psi_{n,q} \eta_{p}(\bx_0) \eta_{q}(\bx_0) \nonumber \\
& = & 2 \int_a^b g(\lambda,\lambda_f) \, \mathrm{d \mathcal{E}_{\zeta,\zeta}(\lambda)}. \label{Eqn:Equiv:Rho}
\end{eqnarray}
where $\zeta = \sum_{p=1}^{N_d} \eta_p(\bx_0) \eta_p$.

\subsubsection{Band structure energy} \label{SubSubSec:BandStructureEnergy}
The band structure energy may be expressed as
\begin{eqnarray}
U_{band} & = & 2 \sum_{n=1}^{N_d} \lambda_n g(\lambda_n,\lambda_f) \nonumber \\
& = & 2 \sum_{p=1}^{N_d} \sum_{n=1}^{N_d} \lambda_n g(\lambda_n,\lambda_f) |\psi_{n,p}|^{2} \nonumber \\
& = & 2 \sum_{p=1}^{N_d} \int_{a}^{b} \lambda g(\lambda,\lambda_f) \, \mathrm{d\mathcal{E}_{\eta_p,\eta_p}}. \label{Eqn:Equiv:BandEnergy}
\end{eqnarray}

\subsubsection{Entropy} \label{SubSubSec:Entropy}
Finally, the entropy follows as
\begin{eqnarray}
S & = & -2 k_B \sum_{n=1}^{N_d} [ g(\lambda_n,\lambda_f) \log g(\lambda_n,\lambda_f) + (1-g(\lambda_n,\lambda_f)) \log (1-g(\lambda_n,\lambda_f))] \nonumber \\
& = & -2 k_B \sum_{p=1}^{N_d} \sum_{n=1}^{N_d} [ g(\lambda_n,\lambda_f) \log g(\lambda_n,\lambda_f) + (1-g(\lambda_n,\lambda_f)) \log (1-g(\lambda_n,\lambda_f)) ] |\psi_{n,p}|^{2} \nonumber \\
& = & -2 k_B \sum_{p=1}^{N_d} \int_{a}^{b} [g(\lambda,\lambda_f) \log g(\lambda,\lambda_f) + (1-g(\lambda,\lambda_f)) \log (1-g(\lambda,\lambda_f))] \, \mathrm{d\mathcal{E}_{\eta_p,\eta_p}}.
\end{eqnarray}

\subsection{Formulation} \label{Sec:LSSGQ:Formulation}
We proceed to describe the proposed LSSGQ method. We solve the nonlinear Kohn-Sham eigenvalue problem using the SCF method (cf., e.~g., \cite{Martin2004}). Customarily, in each iteration of the SCF method, the eigenvalues and eigenfunctions of the Hamiltonian ($\mathcal{H}$) are evaluated for a given electron density. The updated electron density is evaluated using Eqn.~\ref{Eqn:ElectronDensity}, and the process is repeated until convergence. However, we circumvent the evaluation of the eigenfunctions and directly evaluate the electron density by employing the procedure described below.

We seek to evaluate the integral representations of the Fermi energy and electron density described earlier using Guass quadratures.  However, only the operator $\mathcal{H}$ is known and its resolution of the identity $\mathcal{E}(\lambda)$ is unknown {\sl a priori}. In order to overcome this difficulty, we use the spectral theorem to rewrite the recurrence relation given by Eqn. (\ref{Eqn:RecurrenceRelation}) as the Lanczos-type iteration
\begin{eqnarray} \label{Eqn:RecurrenceRelationFunctions}
b_{k+1} v_{k+1} = (\mathcal{H} -a_{k+1}) v_k - b_k v_{k-1} \,, \quad k=0, 1, \ldots, K-1 \nonumber \\
v_{-1} = 0 \,, \quad v_{0} = \zeta \,, \quad b_0 = 1 \label{Eqn:RecurrenceRelationLanczos}
\end{eqnarray}
where
\begin{equation}
a_{k+1} = (\mathcal{H} v_k, v_k) \,, \quad k=0, 1, \ldots, K-1
\end{equation}
and $b_k$ is computed such that $\norm{v_k}=1 \,, \, k=0, 1, \ldots, K-1$. Note that the initial condition $v_0=\zeta$ is chosen depending on the measure $\mathcal{E}_{\zeta,\zeta}(\lambda)$ with respect to which integration needs to be performed. Thereafter, the nodes and corresponding weights of the quadrature rule are ascertained following the procedure described in Section \ref{Sec:LSSGQ:GaussQuadrature}. Therefore, we arrive at
\begin{eqnarray}
N_e & = & 2 \sum_{p=1}^{N_d} \int_a^b g(\lambda, \lambda_{f}) \, \mathrm{d\mathcal{E}_{\eta_p,\eta_p}(\lambda)} \approx 2 \sum_{p=1}^{N_d} \sum_{k=1}^{K} w_{k}^{\eta_p} g(\lambda_k^{\eta_p},\lambda_f) \label{Eqn:LSSGQ:Fermi_level:GQ} \\
\rho(\bx_0) & = &  2 \int_a^b g(\lambda,\lambda_f) \, \mathrm{d\mathcal{E}_{\zeta,\zeta}(\lambda) }\approx 2 \sum_{k=1}^{K} w_k^{\zeta} g(\lambda_k^{\zeta},\lambda_f) \label{Eqn:LSSGQ:Electron_density:GQ}
\end{eqnarray}
where $\zeta = \sum_{p=1}^{N_d} \eta_p(\bx_0) \eta_p$.

Once the self-consistent solution of the Kohn-Sham problem for a given position of the nuclei is calculated, the free energy (Eqn.~\ref{Eqn:FreeEnergy}) is calculated by using Gauss quadrature for the evaluation of the band structure energy and entropy
\begin{eqnarray}
U_{band} & = & 2 \sum_{p=1}^{N_d} \int_{a}^{b} \lambda g(\lambda,\lambda_f) \, \mathrm{d\mathcal{E}_{\eta_p,\eta_p}(\lambda)} \approx 2 \sum_{p=1}^{N_d} \sum_{k=1}^{K} w_{k}^{\eta_p} \lambda_{k}^{\eta_p} g(\lambda_k^{\eta_p},\lambda_f) \,, \label{Eqn:LSSGQ:Band_energy} \\
S & = & -2 k_B \sum_{p=1}^{N_d} \int_{a}^{b} [ g(\lambda,\lambda_f) \log g(\lambda,\lambda_f) + (1-g(\lambda,\lambda_f)) \log (1-g(\lambda,\lambda_f))] \, \mathrm{d\mathcal{E}_{\eta_p,\eta_p}(\lambda)}  \nonumber \\
& \approx & -2 k_B \sum_{p=1}^{N_d} \sum_{k=1}^{K} w_{k}^{\eta_p} [g(\lambda_k^{\eta_p},\lambda_f) \log g(\lambda_k^{\eta_p},\lambda_f) + (1-g(\lambda_k^{\eta_p},\lambda_f)) \log (1-g(\lambda_k^{\eta_p},\lambda_f))]. \label{Eqn:LSSGQ:Entropy}
\end{eqnarray}
To calculate the ground-state free energy, we need to further minimize the free energy with respect to the positions of the nuclei. To this end, we equilibrate the forces on the nuclei given by Eqn.~\ref{Eqn:Force:Nuclei}. We summarize the LSSGQ method in Algorithm \ref{Algo:LSSGQ}.\newline \newline

\begin{algorithm}[htbp] 
Generate guess for positions of the nuclei ($\bR$) \\
\Repeat(Relaxation of atoms){Energy minimized with respect to positions of atoms}
{
        Calculate charge density of the nuclei ($b$) \\
	Generate guess for electron density ($\rho$) \\
	\Repeat(Self-Consistent loop: SCF ){Convergence of self-consistent iteration}
        {
                  Calculate electrostatic potential ($\phi$) by solving Eqn.~\ref{Eqn:Poisson:ElectrostaticPotential} \\
		  Calculate exchange correlation potential ($V_{xc}$) \\
		  {Calculate spectral Gauss quadrature nodes and weights} \\
                  Calculate Fermi energy ($\lambda_{f}$) using Eqn.~\ref{Eqn:LSSGQ:Fermi_level:GQ} \\
                  Calculate electron density ($\rho$) using Eqn.~\ref{Eqn:LSSGQ:Electron_density:GQ} \\
                  Update the electron density (mixing) \\
        }
        Calculate the forces on the nuclei using Eqn.~\ref{Eqn:Force:Nuclei}\\
}
       Evaluate free energy $\mathcal{F}$  using Eqns. \ref{Eqn:LSSGQ:Band_energy} and \ref{Eqn:LSSGQ:Entropy} for the band structure energy and entropy respectively.
\caption{LSSGQ method}
\label{Algo:LSSGQ}
\end{algorithm}

\subsection{Scaling and Performance} \label{Sec:LSSGQ:ScalingPerformance}

In the LSSGQ method, the Fermi energy, electron density, band structure energy and entropy are evaluated by performing Gauss quadrature on their respective integral representations. The cost of evaluating each integral is determined by the calculation of the matrix $\hat{J}_K$ (Eqn.~\ref{Eqn:MatrixJK}) via the recurrence relation (Eqn.~\ref{Eqn:RecurrenceRelationFunctions}) and the evaluation of its eigenvalues and eigenvectors. {The size of $\hat{J}_K$ in turn is ascertained by the number of nodes used for the quadrature ($K$). We claim that the number of quadrature nodes required for achieving a predifined accuracy is independent of system size based on the following argument in the spirit of \cite{Goedecker}. For definiteness, consider the evaluation of the electron density where the function to be integrated is $g(\lambda,\lambda_f)$. Since there are $K$ quadrature nodes, the resolution offered is roughly proportional to $1/K$. Therefore, we can roughly estimate}
\begin{equation}\label{Eqn:K:Scaling}
K \propto \frac{\lambda_{N_d}-\lambda_1}{\triangle \lambda} \,,
\end{equation} 
{where $\triangle \lambda$ is the width of the region where $g(\lambda,\lambda_f)$ takes fractional values. The ratio in Eqn. \ref{Eqn:K:Scaling} is independent of system size, since the number of basis functions required per atom is usually independent of system size. } Further, if the orthonormal basis used to discretize $H$ is localized, then the matrix representation of $\mathcal{H}$ is sparse. In such a situation, the evaluation of the quadrature nodes and weights is independent of system size.  Since we have to evaluate $\mathcal{O}(N_d)$ such quadrature rules, and since $N_d$ scales as number of atoms $M$, we conclude that the entire method scales as  $\mathcal{O}(M)$. This is indeed verified by the results presented in Section \ref{Sec:LSSGQ:NumericalExamplesValidation}.

To understand the performance of the LSSGQ method, consider any integral of the form given by Eqn.~\ref{Eqn:SampleIntegral}. The error incurred by using Gauss quadrature is given by
\begin{equation} \label{Eqn:QuadratureError:Form2}
R[f] = \norm{p_K}_{\zeta}^{2} \frac{f^{2K}(c)}{(2K)!}
\end{equation}
for some $c \in [a, b]$. It is to be expected that the accuracy of the LSSGQ method improves with increasing the temperature $\theta$. Further,  the LSSGQ method is also expected to have spectral convergence with respect to the number of quadrature points ($K$). These properties are verified through the test cases presented in Section \ref{Sec:LSSGQ:NumericalExamplesValidation}.

{In this work, we have developed quadrature rules of degree $2K-1$ by treating both the nodes and weights as unknowns, thus performing Gauss quadrature. It is also possible to fix the positions of the nodes and treat only the weights as unknowns. However, this will result in a quadrature rule of degree $K-1$. Therefore, our choice of Gauss quadrature is optimal in the sense that for a given number of nodes, it has the highest degree quadrature rule. Also, in the proposed approach, the Fermi energy is not needed as an input, which highlights the efficiency of the method.}

\subsection{Finite-difference approximation}\label{Sec:LSSGQ:FD}

{ The method developed above is general and can be implemented with any localized orthogonal basis. We now briefly describe the implementation with the finite-difference approximation.  This provides a concrete example of the method, and also prepares for the examples that follow. Consider a domain in dimension $D$, and discretize it with a uniform finite-difference grid with nodes $\{\bx^p\}_{p=1}^{N_d}$ and spacing $h$.}  By discretizing the Laplacian in the Hamiltonian using finite-differences of chosen order, we obtain the matrix eigenvalue problem
\begin{equation}
{\bf H} \boldsymbol{\psi}_n = \lambda_n \boldsymbol{\psi}_n \,, \quad n = 1, 2, \ldots, N_d
\end{equation}
where $\boldsymbol{\psi}_n = [\psi_{n,1}, \psi_{n,2}, \ldots, \psi_{n,N_d}]^{T}$ is the $n^{th}$ eigenvector or orbital and $\psi_{n,p}$ is the value of the orbital at note $p$. The Lanczos iteration (Eqn.~\ref{Eqn:RecurrenceRelationFunctions}) takes the form
\begin{eqnarray}
b_{k+1}{\bf v}_{k+1} & = & ({\bf H}-a_{k+1}) {\bf v}_k - b_k {\bf v}_{k-1} \,, \,\, k=0, 1, \ldots , K-1 \nonumber \\
{\bf v}_{-1} & = & \begin{pmatrix} 0 \\ \vdots \\ 0 \\ \vdots \\ 0 \end{pmatrix} \,, \quad {\bf v}_{0} = \begin{pmatrix} 0 \\ \vdots \\ 1 \\ \vdots \\0\end{pmatrix} \,, \quad b_0=1 \label{Eqn:LanczosIteration:FD}
\end{eqnarray}
where
\begin{equation}
a_{k+1} = {\bf v}_k {\bf H} {\bf v}_k \,, \,\,\, k=0,1,\ldots,K-1
\end{equation}
and $b_k$ is computed such that $\norm{{\bf v}_k}_{l^2} = 1$. For the $p^{th}$ finite-difference node, only the $p^{th}$ component of the initial vector ${\bf v}_0$ is nonzero and is set to unity. The calculation of the spectral quadrature nodes $\{ \lambda_k^p \}_{k=1}^{K}$ and weights $\{ w_k^{p} \}_{k=1}^{K}$ for each node $p \in [1,N_d]$ proceeds as described in Section \ref{Sec:LSSGQ:Formulation}.  {  Importantly, note that these values may be computed independently at each node}.   Once the spectral quadrature nodes and weights are known, we evaluate the Fermi energy by solving for the constraint
\begin{equation}
N_e = h^D \sum_{p=1}^{N_d} \rho^{p} \label{Eqn:FD:FermiEnergy} \\
\end{equation}
where
\begin{equation}
\rho^{p} = \frac{2}{h^D} \sum_{k=1}^{K} w_k^p g(\lambda_k^p,\lambda_f)  \label{Eqn:FD:ElectronDensity}
\end{equation}
is the electron density at the $p^{th}$ finite-difference node. Furthermore, the band structure energy and entropy can be evaluated
\begin{eqnarray}
U_{band} & = & \sum_{p=1}^{N_d} u^p \label{Eqn:FD:BandEnergy} \,, \\
S & = & \sum_{p=1}^{N_d} s^p \label{Eqn:FD:Entropy}
\end{eqnarray}
where
\begin{eqnarray}
u^p & = &  2 \sum_{k=1}^{K} w_k^p \lambda_k^{p} g(\lambda_k^p,\lambda_f) \,, \label{Eqn:up}\\
s^p & = & -2 k_B \sum_{k=1}^{K} w_{k}^{p} [g(\lambda_k^{p},\lambda_f) \log g(\lambda_k^{p},\lambda_f) + (1-g(\lambda_k^{p},\lambda_f)) \log (1-g(\lambda_k^{p},\lambda_f))]  \label{Eqn:sp}.
\end{eqnarray}
The above equations allow us to intepret $u^p$ and $s^p$ as the nodal band structure energy and nodal entropy.

\subsection{Numerical Examples and Validation} \label{Sec:LSSGQ:NumericalExamplesValidation}

In this section, we proceed to verify the LSSGQ method through selected examples. Our first example concerns a one-dimensional model problem proposed by \cite{Garcia2009}. Since the problem shares many of the features of the linearized Kohn-Sham problem, it provides a convenient test of the accuracy and performance of the LSSGQ method. In our second example, we apply the LSSGQ method to the nonlinear three-dimensional Kohn-Sham problem and evaluate selected crystal properties. Finally, we showcase the scope and versatility of the LSSGQ method by studying the phenomenon of surface relaxation.

\subsubsection{One-dimensional model problem} \label{Sec:LSSGQ:Numerical:1d}

Consider a one-dimensional chain of atoms positioned with unit spacing at $\{ R_J \}_{J=1}^{M}$. Let the effective potential due to an atom at $R_J$ be given by \citep{Garcia2009}
\begin{equation}
V_J(x) = -\frac{\alpha}{\sqrt{2 \pi \beta^2}}\exp{\left(-\frac{(x-R_J)^2}{2\beta^2}\right)}
\end{equation}
and therefore the total potential at any point is given by $V(x)=\sum_{J=1}^{M}V_J(x)$. Here, $\alpha$ and $\beta$ represent the depth and width of the wells respectively. The ground-state properties are determined through the eigenvalue problem
\begin{equation}
\mathcal{H} \psi_n = \lambda_n \psi_n \,, \quad n = 1, 2 \ldots
\end{equation}
where the Hamiltonian is given by
\begin{equation}
\mathcal{H} = -\frac{1}{2} \frac{d^2}{d x^2} + V(x).
\end{equation}
The Fermi energy, electron density and energy are calculated using the relations
\begin{eqnarray}
N_e & = & \sum_{n=1}^{N_d} g(\lambda_n,\lambda_f) \,, \\
\rho(x) & = & \sum_{n=1}^{N_d} g(\lambda_n,\lambda_f) |\psi_n(x)|^2 \,, \\
\mathcal{E}_{\sigma} & = & \sum_{n=1}^{N_d} g(\lambda_n,\lambda_f) \lambda_n.
\end{eqnarray}

We introduce a $12^{th}$ order accurate finite-difference approximation for the Laplacian and use the LSSGQ method described in Section \ref{Sec:LSSGQ:FD}. By appropriately choosing the parameters $\alpha$ and $\beta$, we can vary the band gap of the system, thereby choosing between insulating, semiconducting and metallic systems. Further details can be found in \cite{Garcia2009}. We consider two sets of parameters $(\alpha,\beta)=(100,0.3)$ and $(\alpha,\beta)=(10,0.45)$, which we designate as insulator and metal, respectively. We also study the effect of temperature on the efficacy of the method, by considering two extreme cases corresponding to $\sigma=0.0001$ and  $\sigma = 1.0$. In the results presented below, the error in energy is defined as the difference between the energies obtained by the LSSGQ method and diagonalization. Similarly, error in electron density is the $L^2$ norm of the difference in electron densities obtained on using the LSSGQ method and diagonalization. We note that all errors have been normalized by the corresponding quantities obtained via diagonalization.

\begin{figure}[h!]
\centering
\subfloat[$\sigma=1$] {\label{fig:UvsK_beta1}
\includegraphics[keepaspectratio=true,width=0.48\textwidth]
{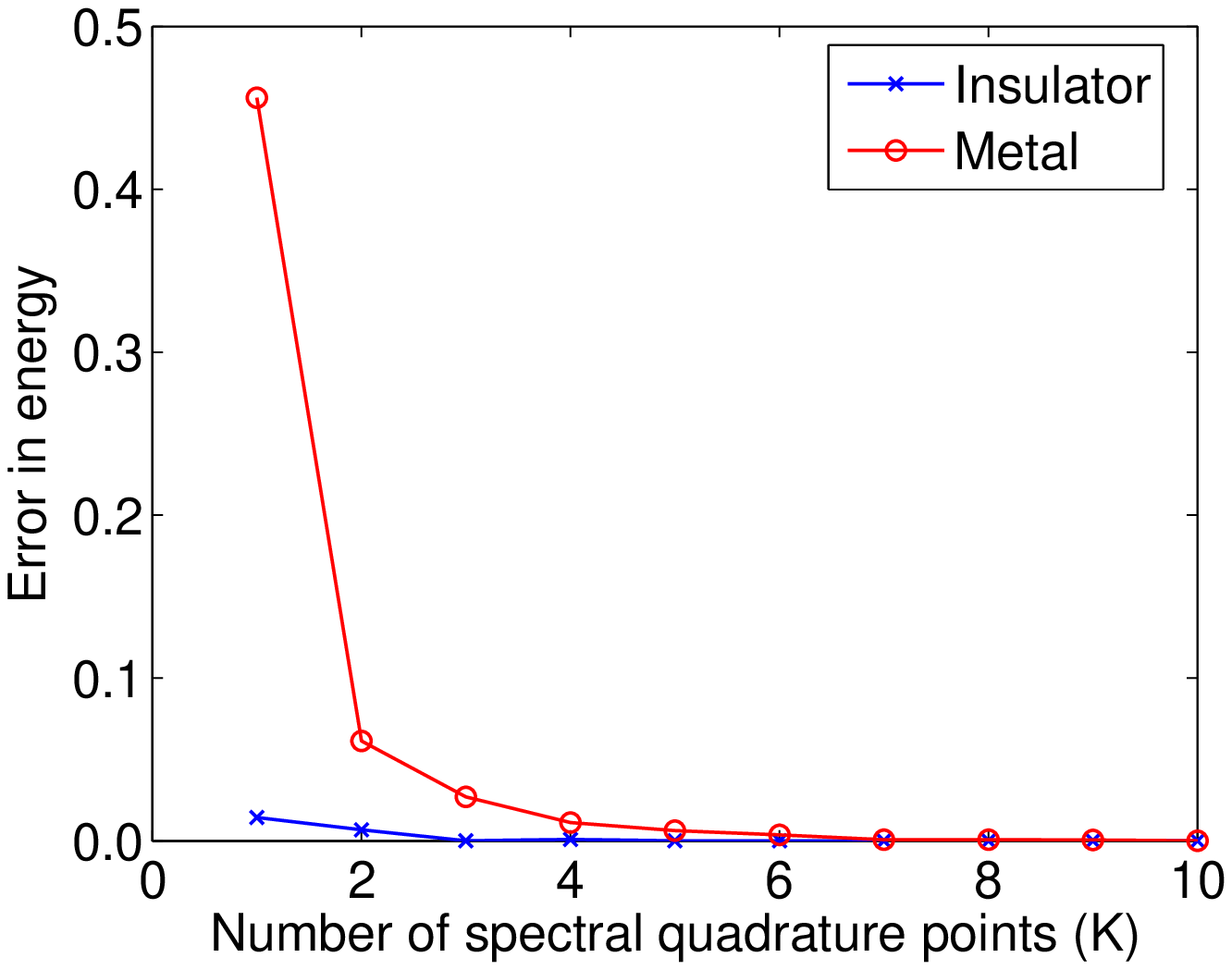}}
\subfloat[$\sigma=0.0001$] {\label{fig:UvsK_beta10000}
\includegraphics[keepaspectratio=true,width=0.48\textwidth]
{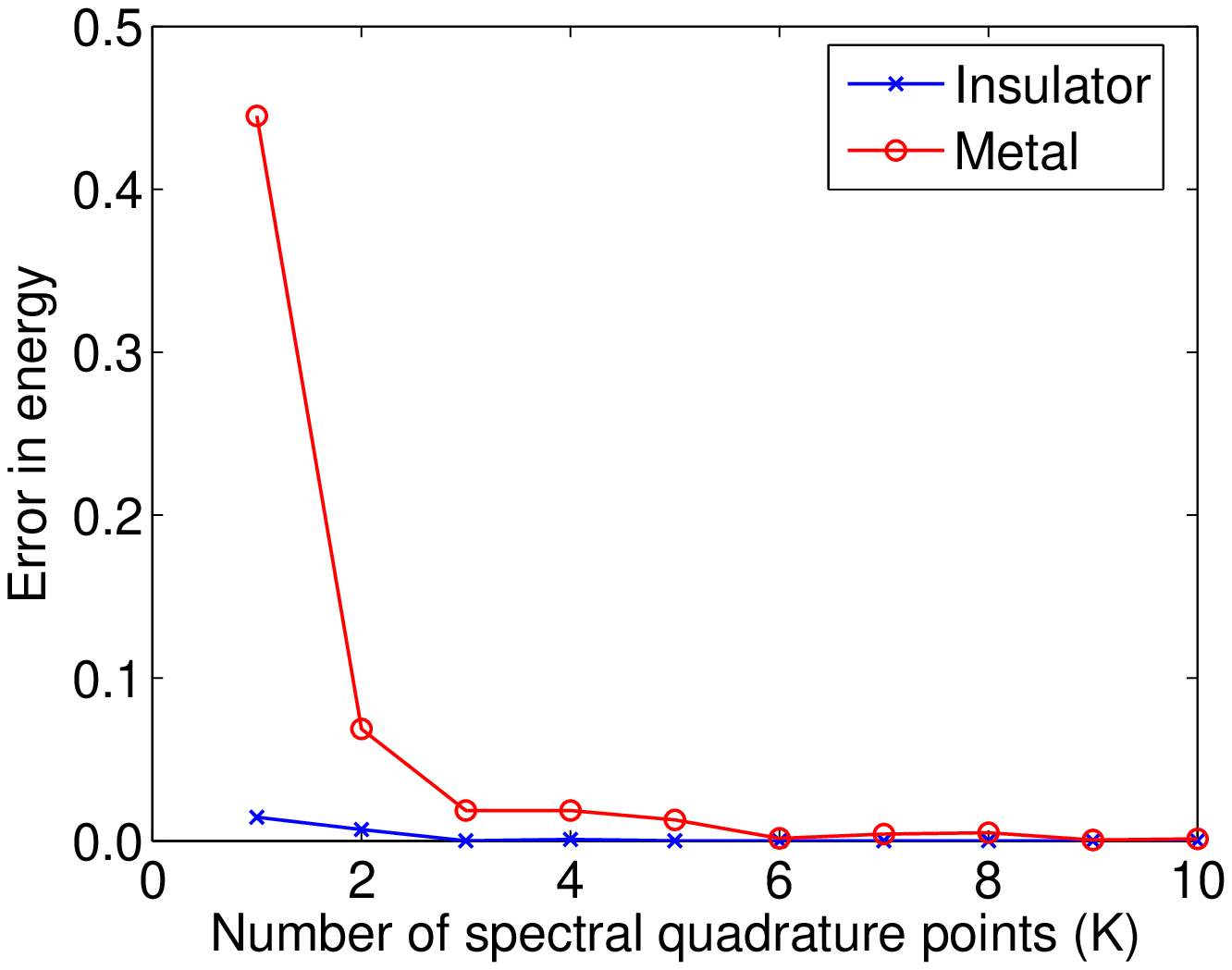}}
\caption{Convergence of the LSSGQ method for the one-dimensional model problem}
\label{Fig:Convergence}
\end{figure}

\paragraph{Non-periodic calculations} 
{For these calculations, we choose a chain of $101$ atoms, i.e. $M=101$.} The convergence of the LSSGQ method with respect to the number of spectral quadrature points for both metals and insulators is presented in Fig.~\ref{Fig:Convergence}. The rapid convergence of the energy for both metals and insulators is clear from the figure. For insulators, the convergence is independent of temperature, a consequence of the presence of a band gap. By contrast, convergence for metals is significantly accelerated on increasing the temperature.

\begin{figure}[h!]
\centering
\subfloat[$\sigma=1$] {\label{fig:EdvsK_beta1}
\includegraphics[keepaspectratio=true,width=0.48\textwidth]
{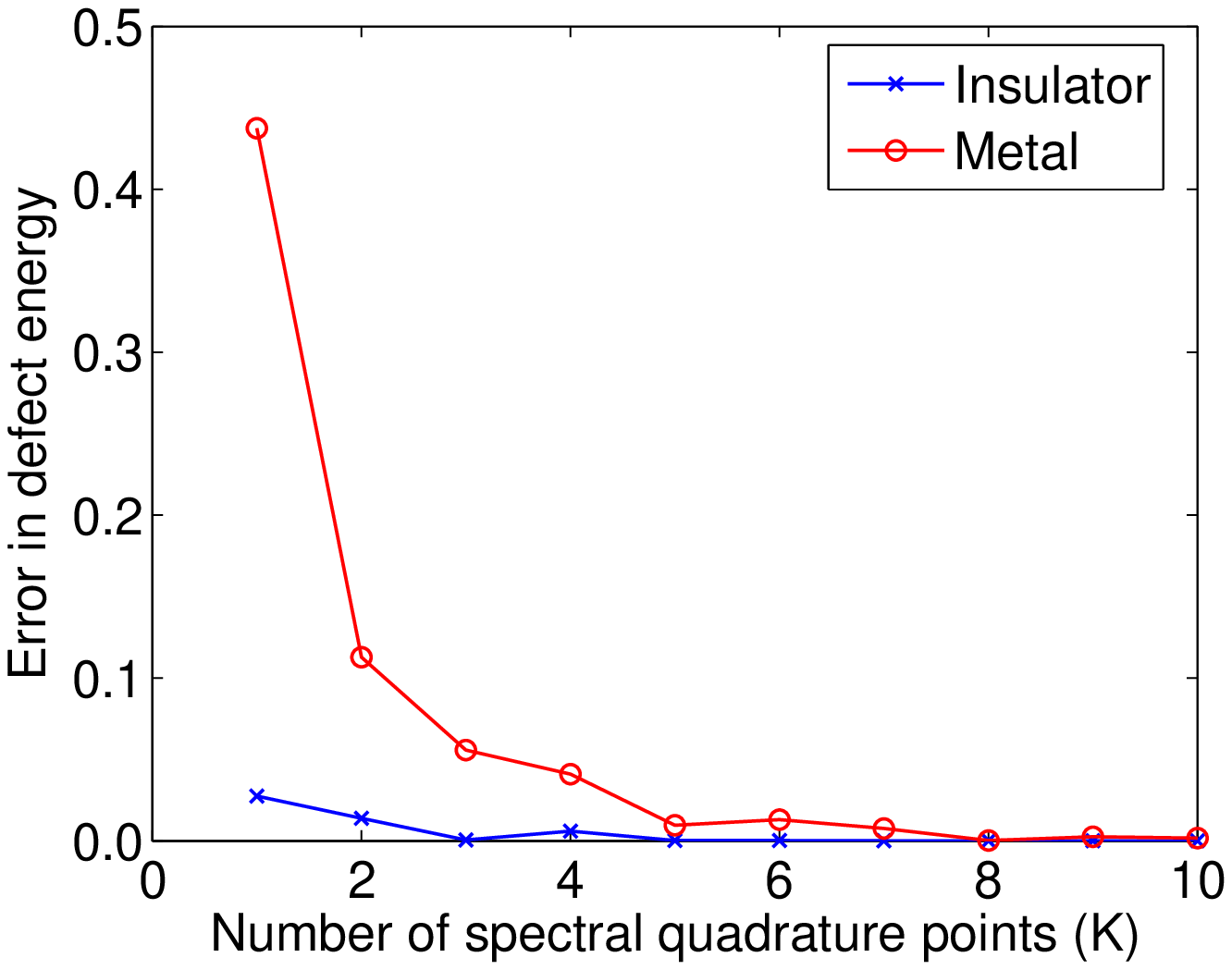}}
\subfloat[$\sigma=0.0001$] {\label{fig:EdvsK_beta10000}
\includegraphics[keepaspectratio=true,width=0.48\textwidth]
{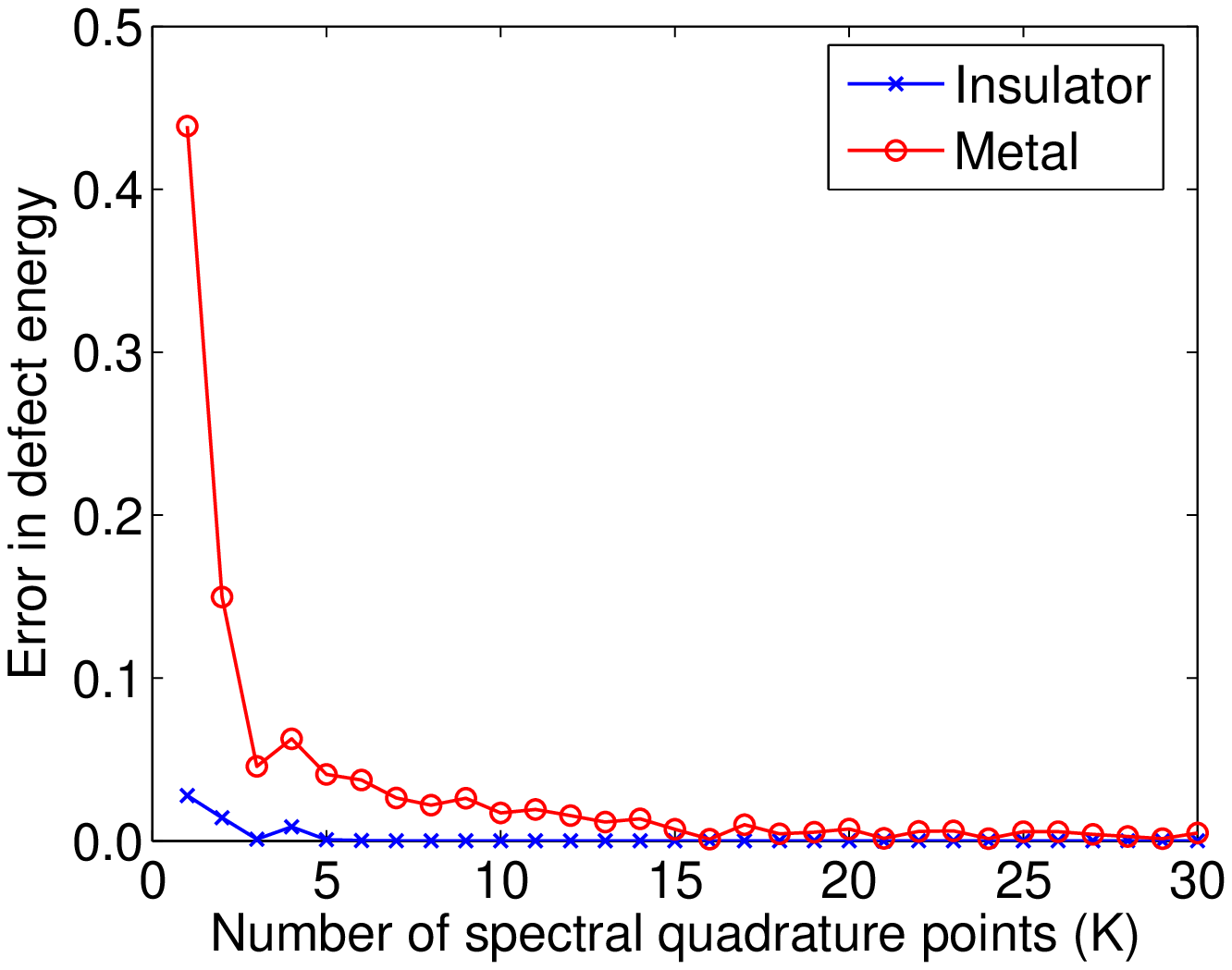}} \\
\subfloat[$\sigma=1$] {\label{fig:err_rhovsK_beta1}
\includegraphics[keepaspectratio=true,width=0.48\textwidth]
{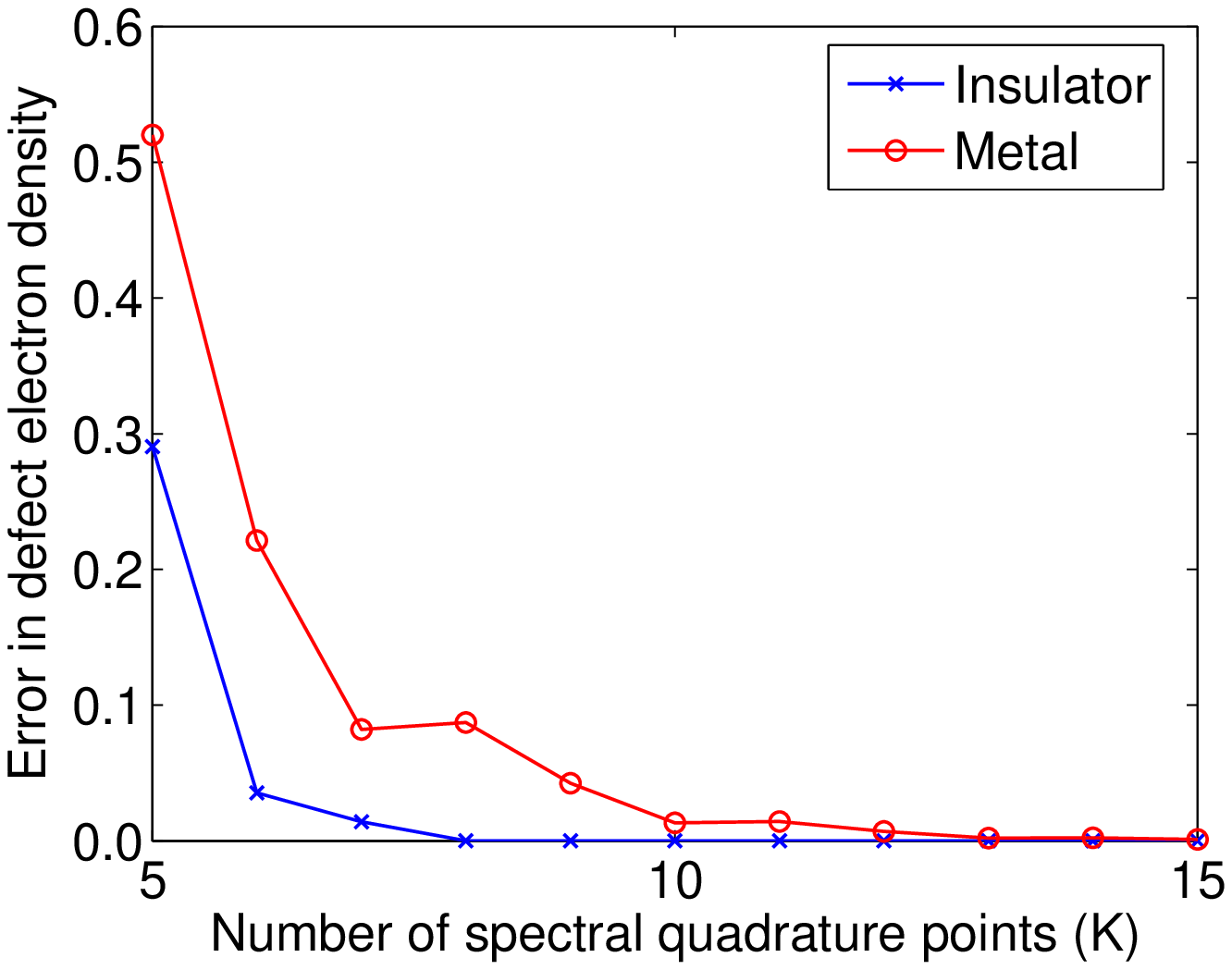}}
\subfloat[$\sigma=0.0001$] {\label{fig:err_rhovsK_beta10000}
\includegraphics[keepaspectratio=true,width=0.48\textwidth]
{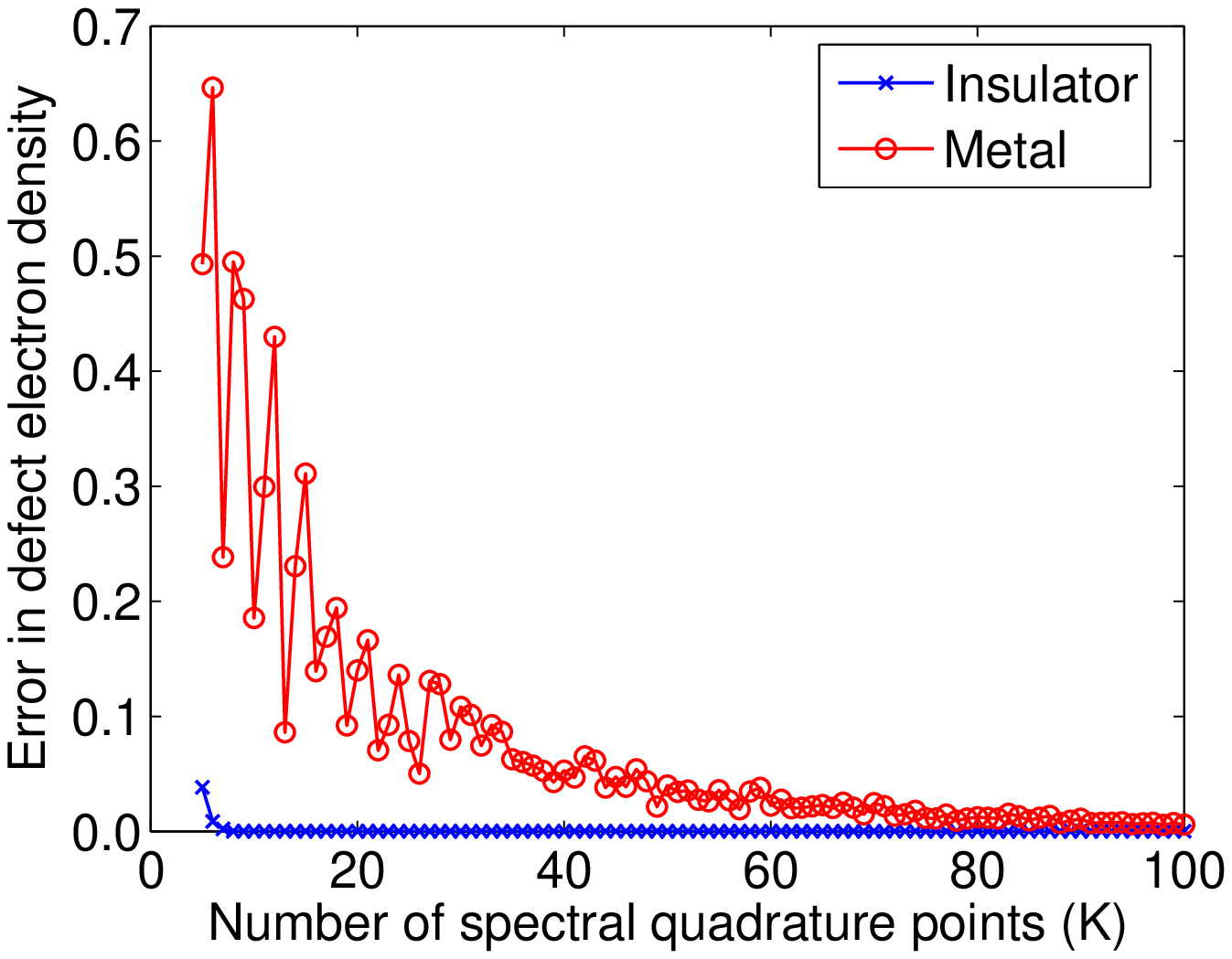}}
 \caption{Convergence of the defect energy and defect electron density for the one-dimensional model problem}
\label{Fig:Convergence:DefectEnergyElectronDensity}
\end{figure}

Next, we introduce a defect by removing the center atom and evaluate the defect energy as well as the defect electron density. The defect energy and defect electron density are defined as the difference in energies and electron densities between systems with and without the defect. Accurately predicting these quantities is challenging, as it requires the calculation of relatively small differences. The results so obtained are presented in Fig.~\ref{Fig:Convergence:DefectEnergyElectronDensity}. For the insulator, there is rapid convergence with number of spectral quadrature points and the convergence is independent of the temperature. However, for the metal we notice that the number of spectral quadrature points required is dependent on the temperature, with fewer points required on increasing the temperature.

\paragraph{Periodic calculations}

\begin{figure}[h!]
\centering
\subfloat[$\sigma=1$] {\label{fig:UpervsK_beta1}
\includegraphics[keepaspectratio=true,width=0.48\textwidth]
{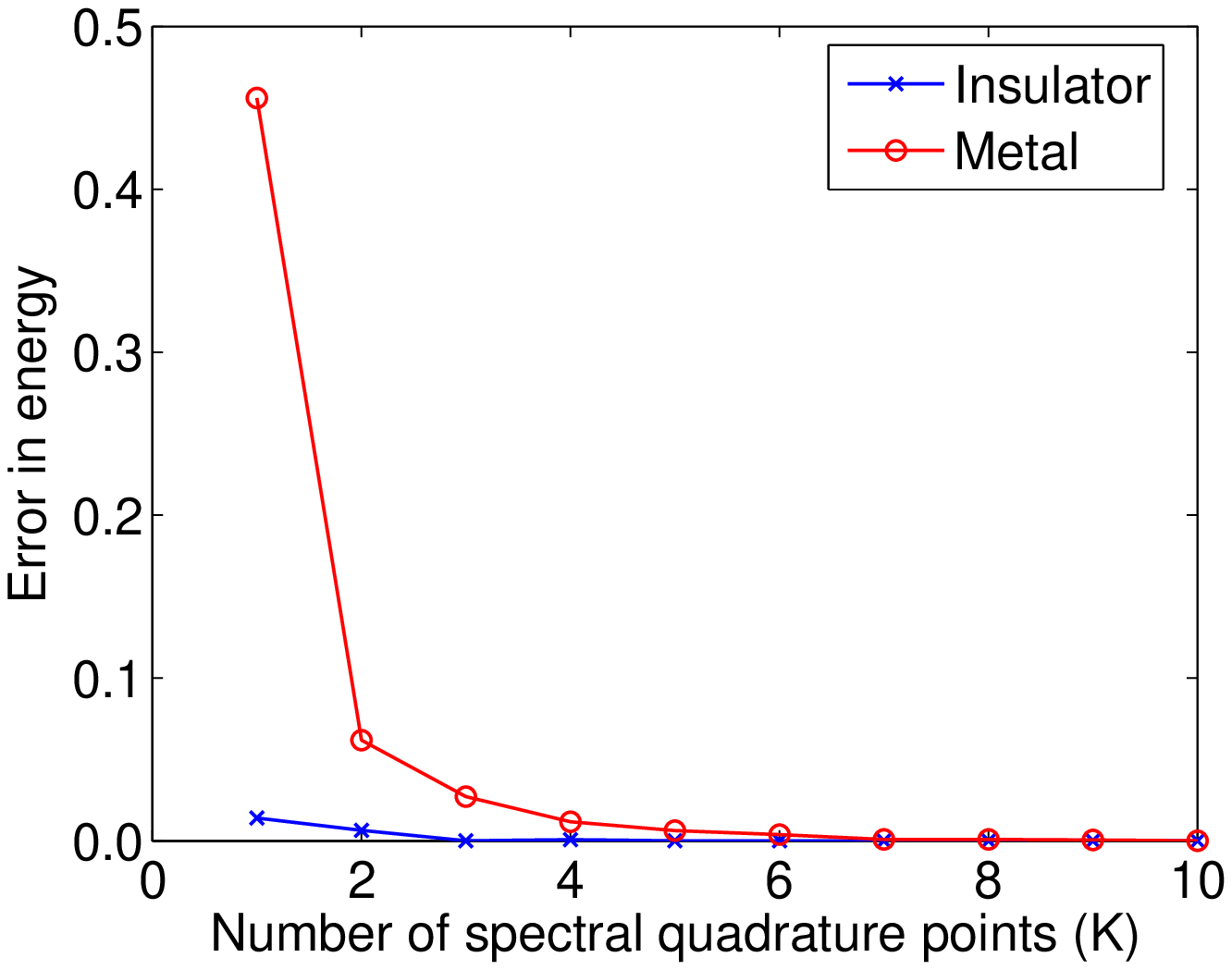}}
\subfloat[$\sigma=0.0001$] {\label{fig:UpervsK_beta10000}
\includegraphics[keepaspectratio=true,width=0.48\textwidth]
{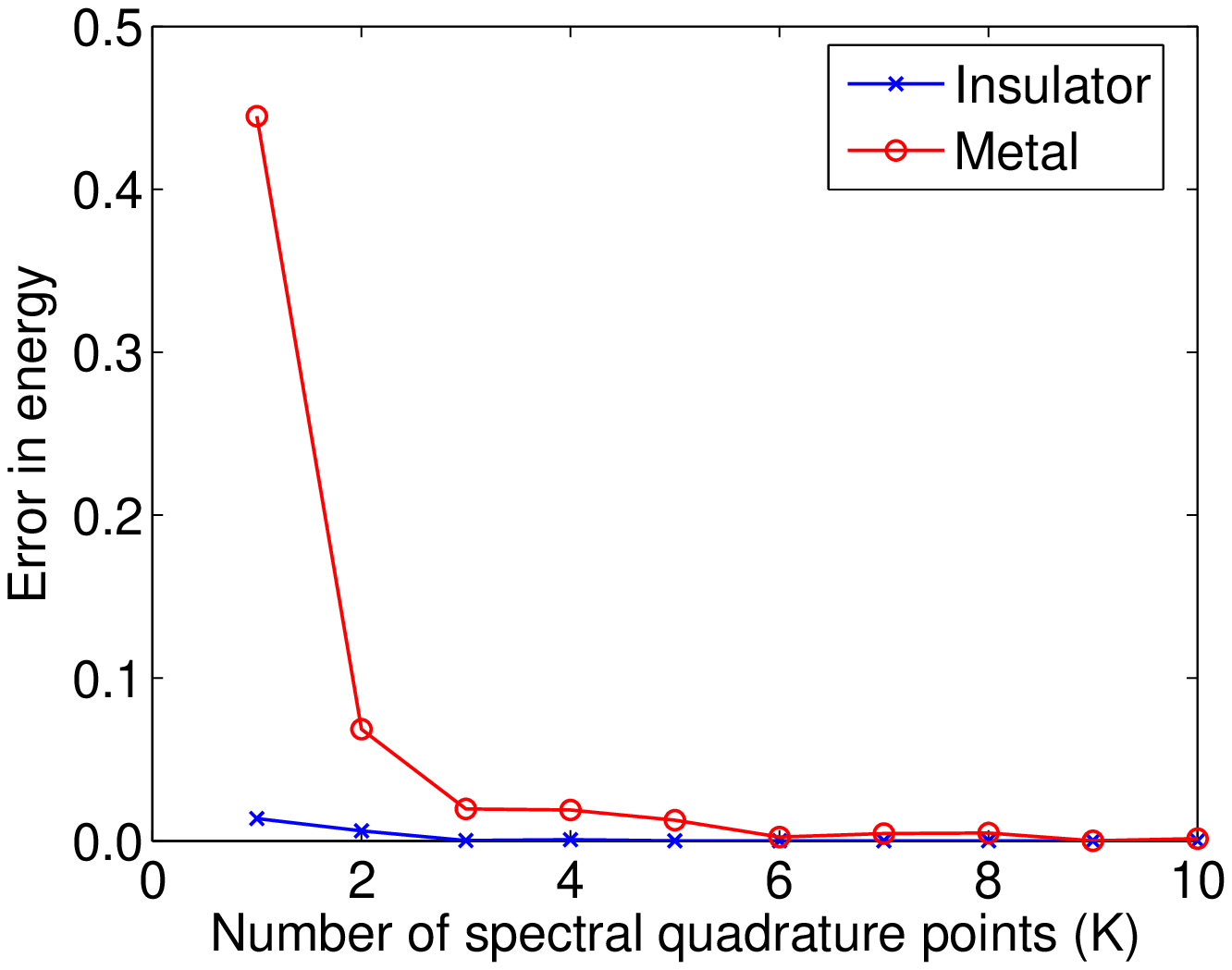}} 
\caption{Convergence of the LSSGQ method for the periodic one-dimensional model problem}
\label{Fig:Convergence:Periodic}
\end{figure}

Consider next an infinite chain of atoms with unit spacing. We define a two atom unit cell as the representative volume $\Omega_{RV}$. Since the potential decays exponentially, we evaluate the potential inside $\Omega_{RV}$ by considering the contribution of atoms which are within a cutoff radius. We then follow the procedure outlined in Appendix \ref{App:LSSGQ:periodic} to evaluate the Fermi energy, electron density and energy per atom. The results so obtained are presented in Fig.~\ref{Fig:Convergence:Periodic}. As observed in the non-periodic calculations, convergence in insulators is independent of temperature, whereas for metals convergence is significantly accelerated with increasing temperature.

\paragraph{Performance and Scaling}

\begin{figure}[h!]
\centering
\subfloat[Insulator] {\label{fig:Insulator_Convergence}
\includegraphics[keepaspectratio=true,width=0.48\textwidth]
{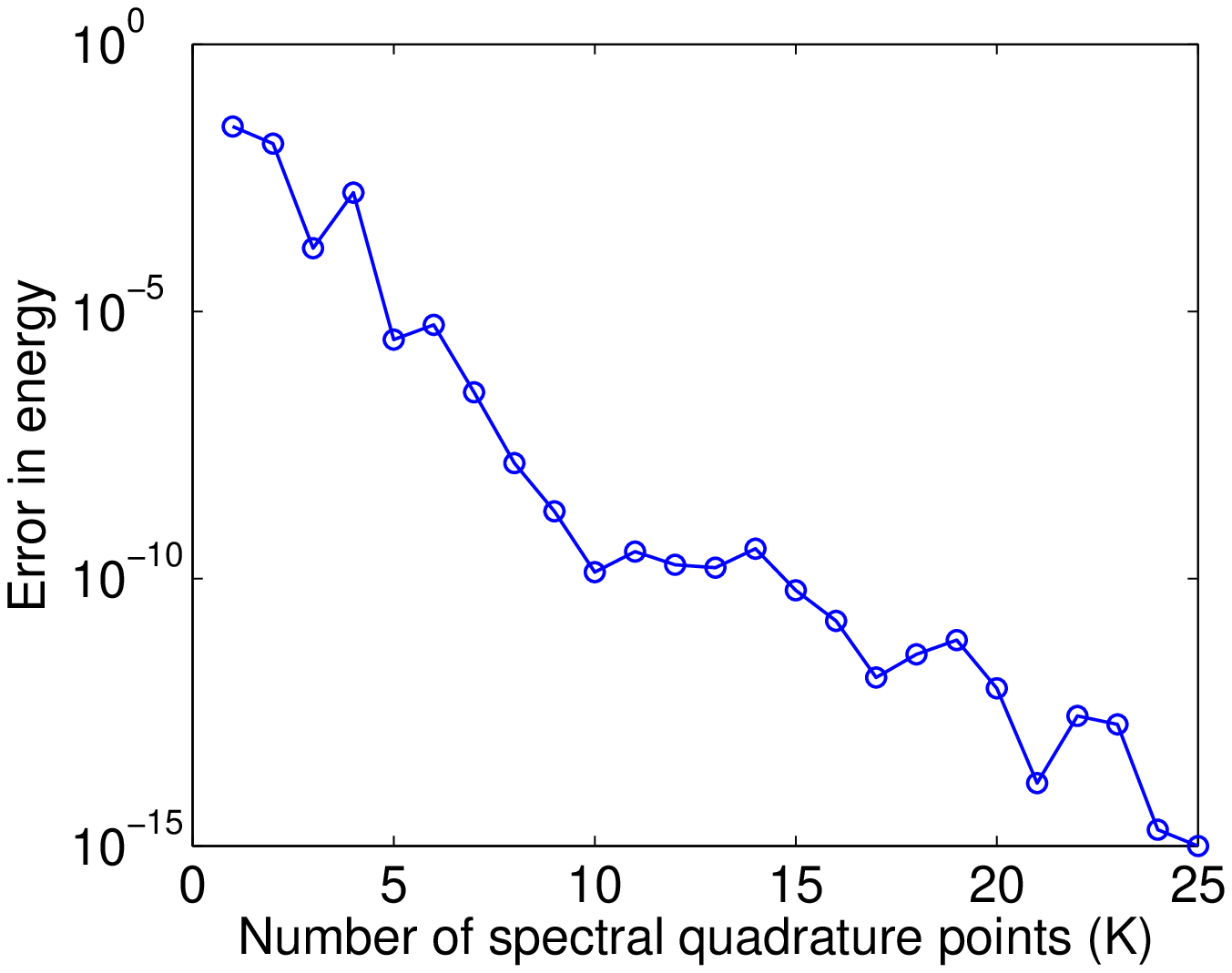}}
\subfloat[Metal] {\label{fig:Conductor_Convergence}
\includegraphics[keepaspectratio=true,width=0.48\textwidth]
{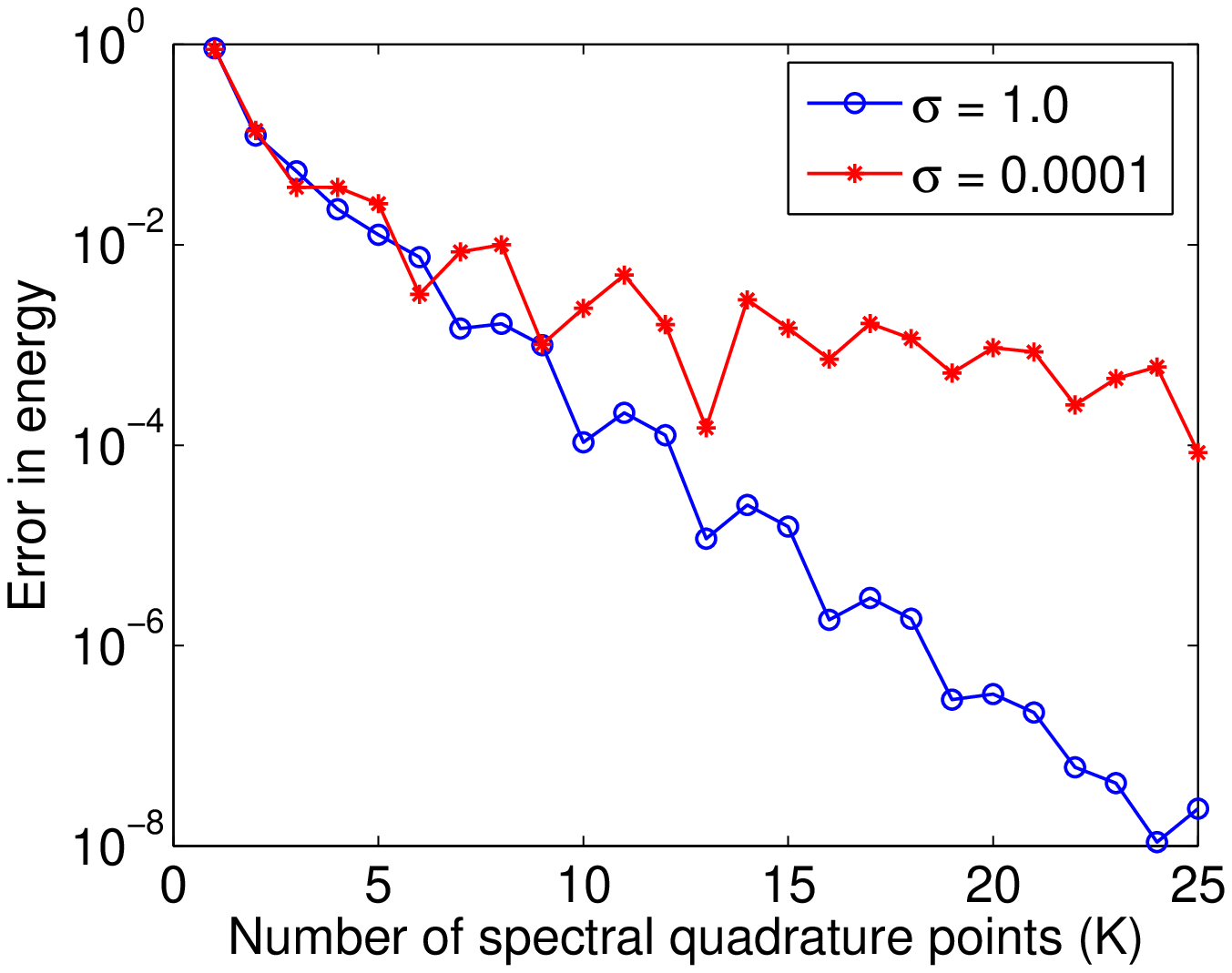}} 
\caption{Spectral convergence of the LSSGQ method}
\label{Fig:Convergence:Insulator_Conductor}
\end{figure}

We now examine the rate of convergence as well as the scaling properties of the LSSGQ method. {First, we perform non-periodic calculations on a chain of $101$ atoms ($M=101$). We plot the convergence in energy with respect to the number of spectral quadrature points in Fig.~\ref{Fig:Convergence:Insulator_Conductor}. It is clear that we get spectral convergence, as is to be expected, since Gauss quadrature is itself a spectrally convergent method. }

\begin{figure}[h!]
\centering
\includegraphics[keepaspectratio=true,width=0.6\textwidth]
{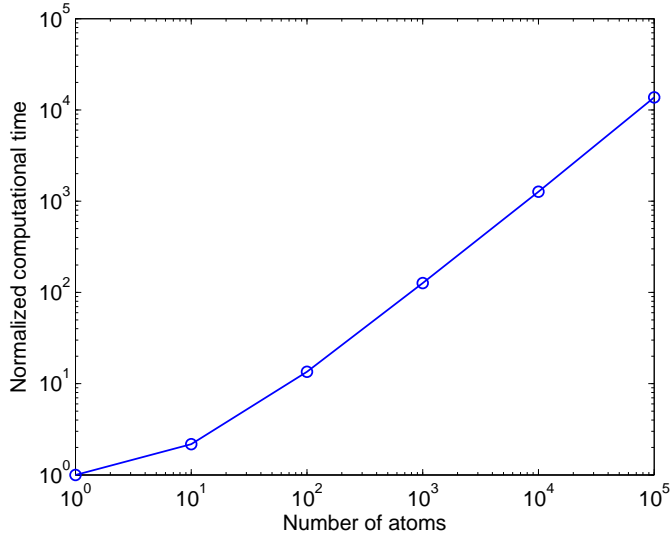}
\caption{Linear-scaling nature of the LSSGQ method}
\label{fig:linearscaling}
\end{figure}

{Next, we study the scaling of the LSSGQ method in terms of the computational time taken versus the system size. We perform non-periodic calculations for metallic systems ranging from $1$ atom to $100,000$ atoms, all on a single computer. This has been showcased in Fig.~\ref{fig:linearscaling}, which highlights that the LSSGQ method is indeed linear-scaling. It should be noted that such large systems are intractable for orbital formulations, especially when solving on a single computer.}

\subsubsection{Kohn-Sham problem} \label{Sec:LSSGQ:Validation:KS}

In the previous section, we have verified the LSSGQ method by means of a one-dimensional chain of atoms interacting through a model potential. In our next example, we apply the LSSGQ method to the solution of the Kohn-Sham problem using the LDA (\cite{Kohn1965}) for the exchange correlation energy, and the `Evanescent Core' pseudopotential approximation \citep{Fiolhais1995}. Specifically, we evaluate the bulk properties of body centered cubic (BCC) crystals of sodium, lithium and study the phenomenon of (001) surface relaxation of BCC sodium.

In our calculations, we utilize the procedure outlined in Algorithm \ref{Algo:LSSGQ} to solve for the ground-state properties. Specifically, we use a $6^{th}$ order accurate finite-difference discretization of the Laplacian. We solve for the equilibrium atomic positions through the BFGS quasi-Newton method with a cubic line search procedure (cf., e.~g., \cite{Curtis2002}). For the SCF method, we use the generalized Broyden method \citep{Fang2009} for convergence acceleration. We handle the periodicity in the system for the eigenvalue problem through the procedure outlined in Appendix \ref{App:LSSGQ:periodic}. We calculate the Fermi energy by using a combination of bisection, secant and inverse quadratic interpolation methods \citep{Forsythe1973} to solve Eqn.~\ref{Eqn:LSSGQ:Fermi_level:GQ}. We solve the Poisson equation (Eqn.~\ref{Eqn:Poisson:ElectrostaticPotential}) with the generalized minimal residual method (GMRES - \cite{GMRES}), wherein we evaluate the charge density $b(\bx,\bR)$ by employing a prespecified radius. In order to evaluate the free energy of the system, we need to perform spatial integrations. To this end, we employ the trapezoidal rule, wherein each quantity is assumed to be constant in a cube of side $h$ around each finite-difference point. Here, $h$ denotes the spacing of the finite-difference nodes. Finally, we choose $\sigma=0.8$ eV for all our calculations and extrapolate to find the ground-state energy at absolute zero using Eqn.~\ref{Eqn:FiniteTemperatureExtrapolation}.

\paragraph{Crystal properties}

\begin{figure}[h!]
\centering
\subfloat[Sodium] {\label{fig:Convergence_Na_K}
\includegraphics[keepaspectratio=true,width=0.48\textwidth]
{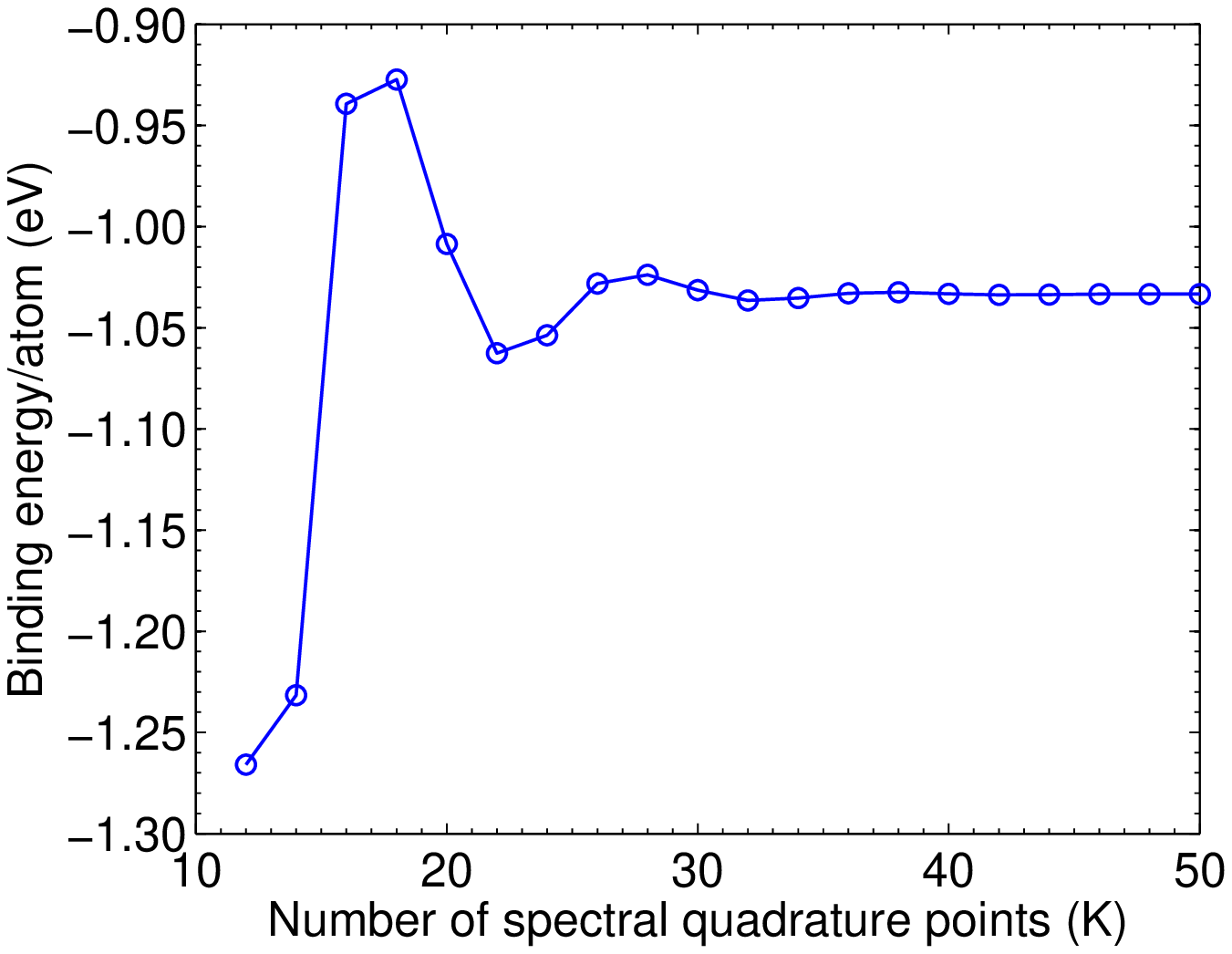}}
\subfloat[Lithium] {\label{fig:Convergence_Li_K}
\includegraphics[keepaspectratio=true,width=0.48\textwidth]
{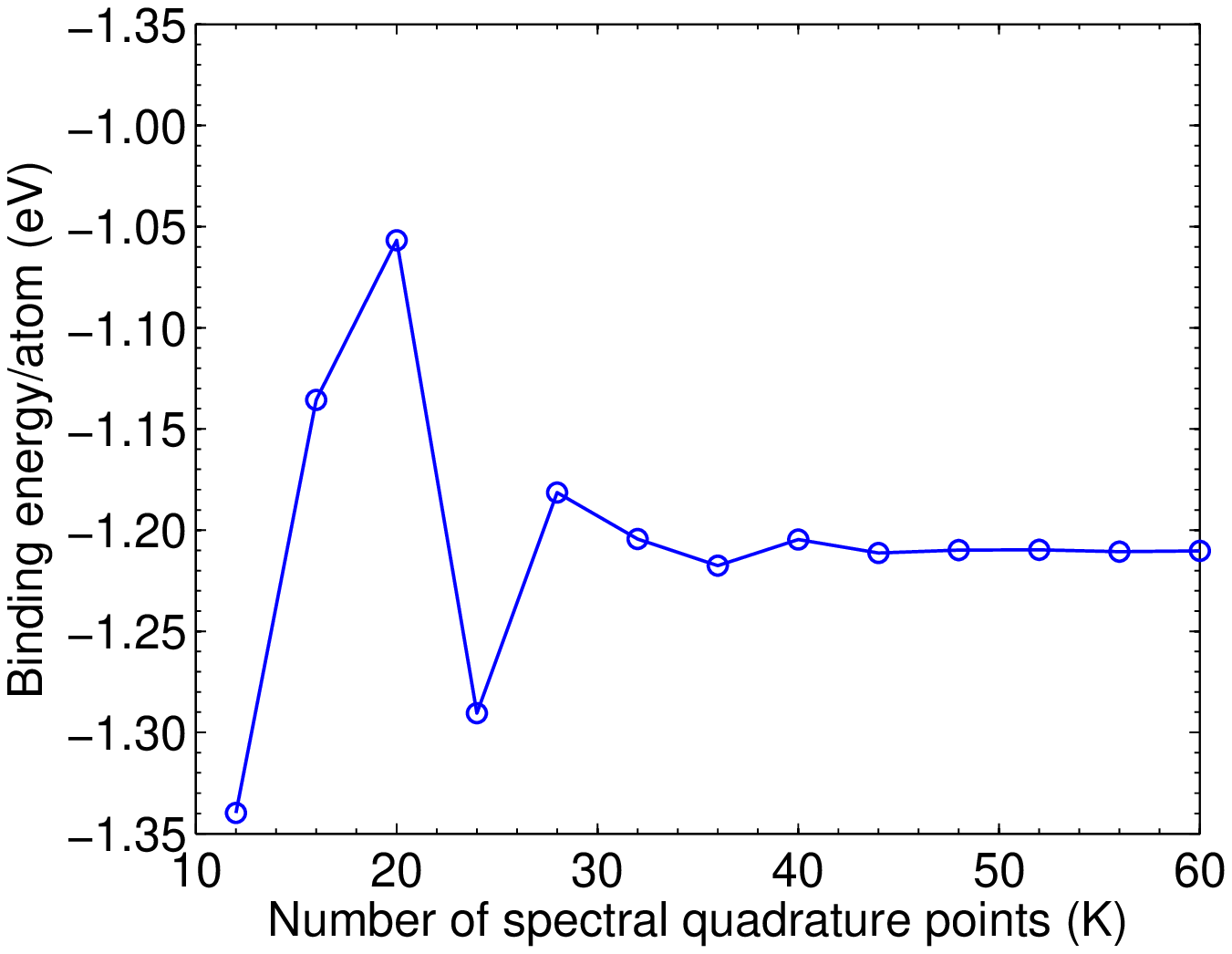}}
\caption{Convergence of the LSSGQ method for evaluating the crystal properties}
\label{Fig:Convergence_Na_Li}
\end{figure}

\begin{figure}[h!]
\centering
\subfloat[Sodium] {\label{fig:BindingEnergy_vs_lattice_constant_Na}
\includegraphics[keepaspectratio=true,width=0.48\textwidth]
{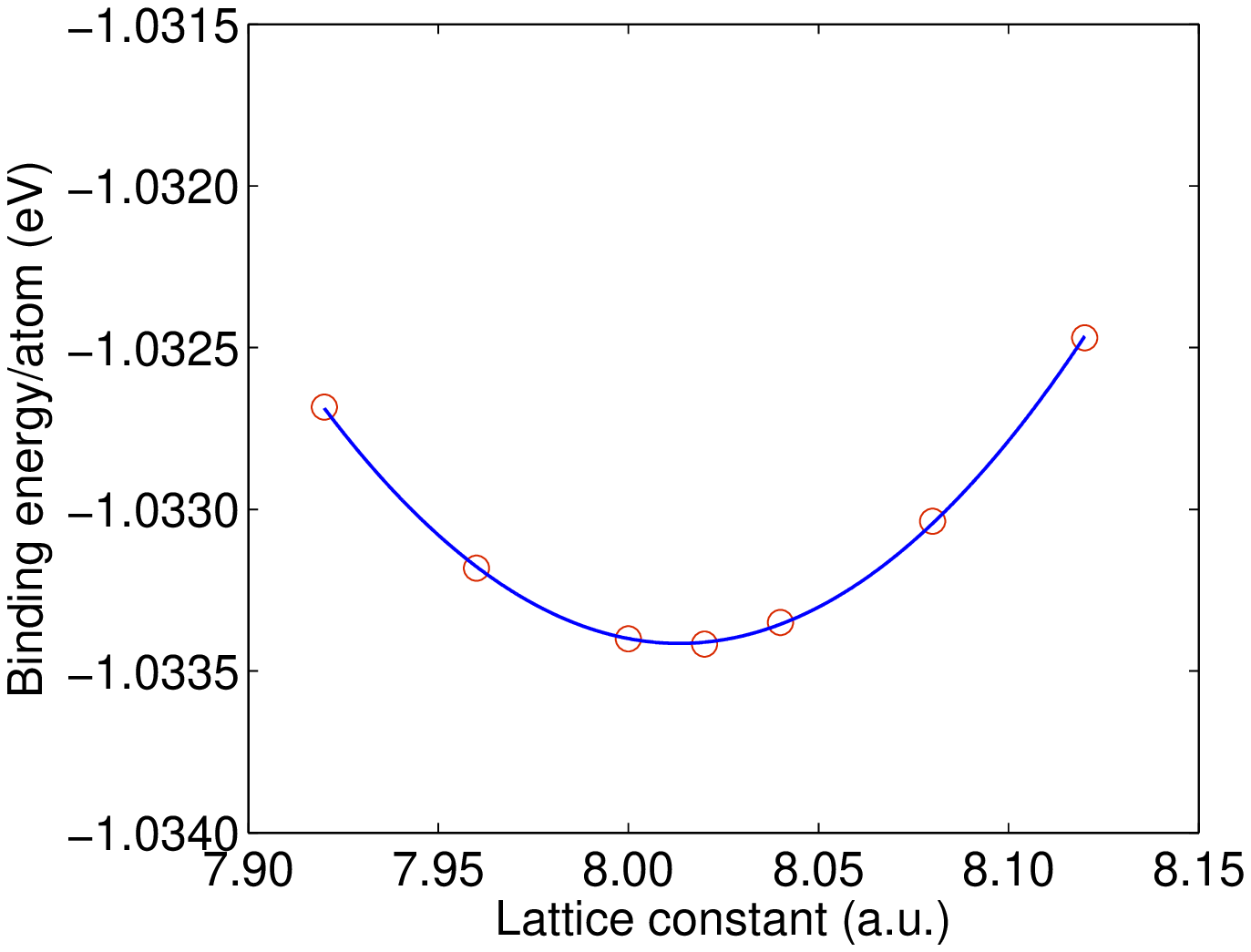}}
\subfloat[Lithium]
{\label{fig:BindingEnergy_vs_lattice_constant_Li}
\includegraphics[keepaspectratio=true,width=0.48\textwidth]
{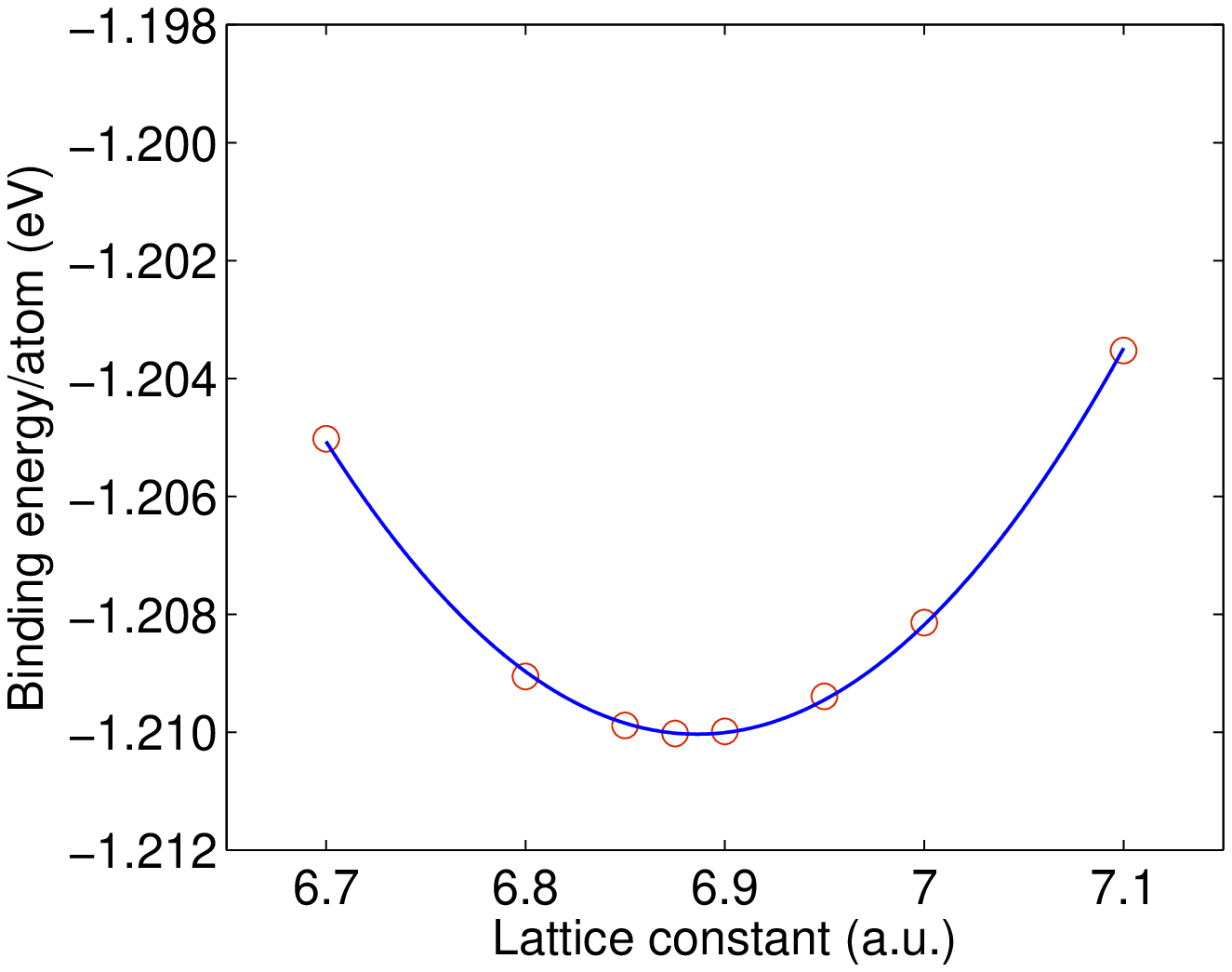}}
\caption{Binding energy as a function of lattice constant}
\label{Fig:BE_a}
\end{figure}

We begin by evaluating the bulk properties of BCC sodium and lithium. We designate the BCC unit cell as the representative volume and use periodic boundary conditions. First, we verify the convergence of the LSSGQ method. To do so, we plot the binding energy per atom as a function of the number of spectral quadrature points ($K$) in Fig.  \ref{Fig:Convergence_Na_Li}. It is clear that we obtain rapid convergence, highlighting the efficacy of the LSSGQ method. Next, we plot the binding energy per atom as a function of lattice constant in Fig.~\ref{Fig:BE_a}. Using a cubic fit to this data, we calculate the cohesive energy, equilibrium lattice constant and bulk modulus. The results so obtained are presented in Tables \ref{table:LSSGQ:periodic:sodium} and \ref{table:LSSGQ:periodic:lithium} with a comparison to previous results obtained by \cite{Fiolhais1995}. There is good agreement in the cohesive energy and the lattice constant. {However, there is a reasonable disagreement in the bulk modulus. The fact that we obtain a smooth curve for the binding energy versus the lattice constant (Fig. \ref{Fig:BE_a}) gives us confidence in our results. Nevertheless, the discrepancy warrants further investigation.} 

\begin{table}[h!]
\begin{center}
\caption{Crystalline properties of BCC sodium}
  \label{table:LSSGQ:periodic:sodium}
\begin{tabular}{cccc}
    \hline
 Property & LSSGQ & \cite{Fiolhais1995} & Experiment \citep{Fiolhais1995} \\
\hline
Cohesive energy (eV/atom) & -1.03 & -1.02 & -1.04 \\
Lattice constant (a.~u.) & 8.01 & 8.21 & 8.21 \\
Bulk modulus (GPa) & 5.0 & 7.1 & 7.3 \\
\hline
\end{tabular}
\end{center}
\end{table}

\begin{table}[h!]
\begin{center}
\caption{Crystalline properties of BCC lithium}
  \label{table:LSSGQ:periodic:lithium}
\begin{tabular}{cccc}
    \hline
 Property & LSSGQ & \cite{Fiolhais1995} & Experiment \citep{Fiolhais1995} \\
\hline
Cohesive energy (eV/atom) & -1.21 & -1.31 & -1.0 \\
Lattice constant (a.~u.) & 6.87 & 6.77 & 6.77 \\
Bulk modulus (GPa) & 10.0 & 14.0 & 13.3 \\
\hline
\end{tabular}
\end{center}
\end{table}

\paragraph{Surface relaxation}

Next we study the (001) surface relaxation of BCC sodium. We choose the representative volume ($\Omega_{RV}$) to be a rectangular cuboid of dimensions $a \times a \times h$, where `$a$' is the lattice constant. Furthermore, $h=h_{vac}+h_{cell}$, where $h_{vac}=n_1 a$ and $h_{cell}=n_2 a$ ($n_1, n_2 \in \mathbb{Z}$) are the heights of vacuum and material included in $\Omega_{RV}$, respectively. We use periodic boundary conditions in the $x_1, x_2$ directions. In addition, we use zero Dirichlet boundary conditions on the top face of $\Omega_{RV}$ and assume that the perfectly periodic solution is attained in the bottommost unit cell of $\Omega_{RV}$. We relax the atoms only along the $x_3$ direction. We evaluate the surface energy using the relation
\begin{equation}
\mathcal{E}_{surface} = \frac{\mathcal{E}_{\Omega_{RV}}-N_e\mathcal{E}_{coh}}{a^2}
\end{equation}
where $\mathcal{E}_{\Omega_{RV}}$ is the total energy of $\Omega_{RV}$, $\mathcal{E}_{coh}$ is the cohesive energy for a perfect crystal and $N_e$ is the number of electrons in $\Omega_{RV}$.

\begin{figure}[h!]
\centering
\includegraphics[keepaspectratio=true,width=0.48\textwidth]
{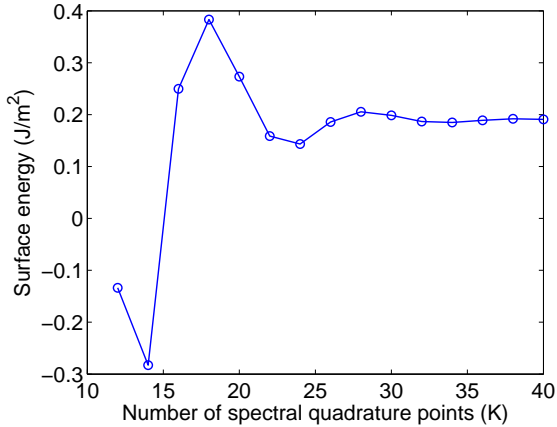}
\caption{Convergence of LSSGQ method for the (001) surface relaxation of BCC sodium}
\label{Fig:Convergence_surface_Na}
\end{figure}

\begin{figure}[h!]
\centering
\subfloat[Mid plane] {\label{Fig:ElectronDensity:Surface:Mid}
\includegraphics[keepaspectratio=true,width=0.50\textwidth]
{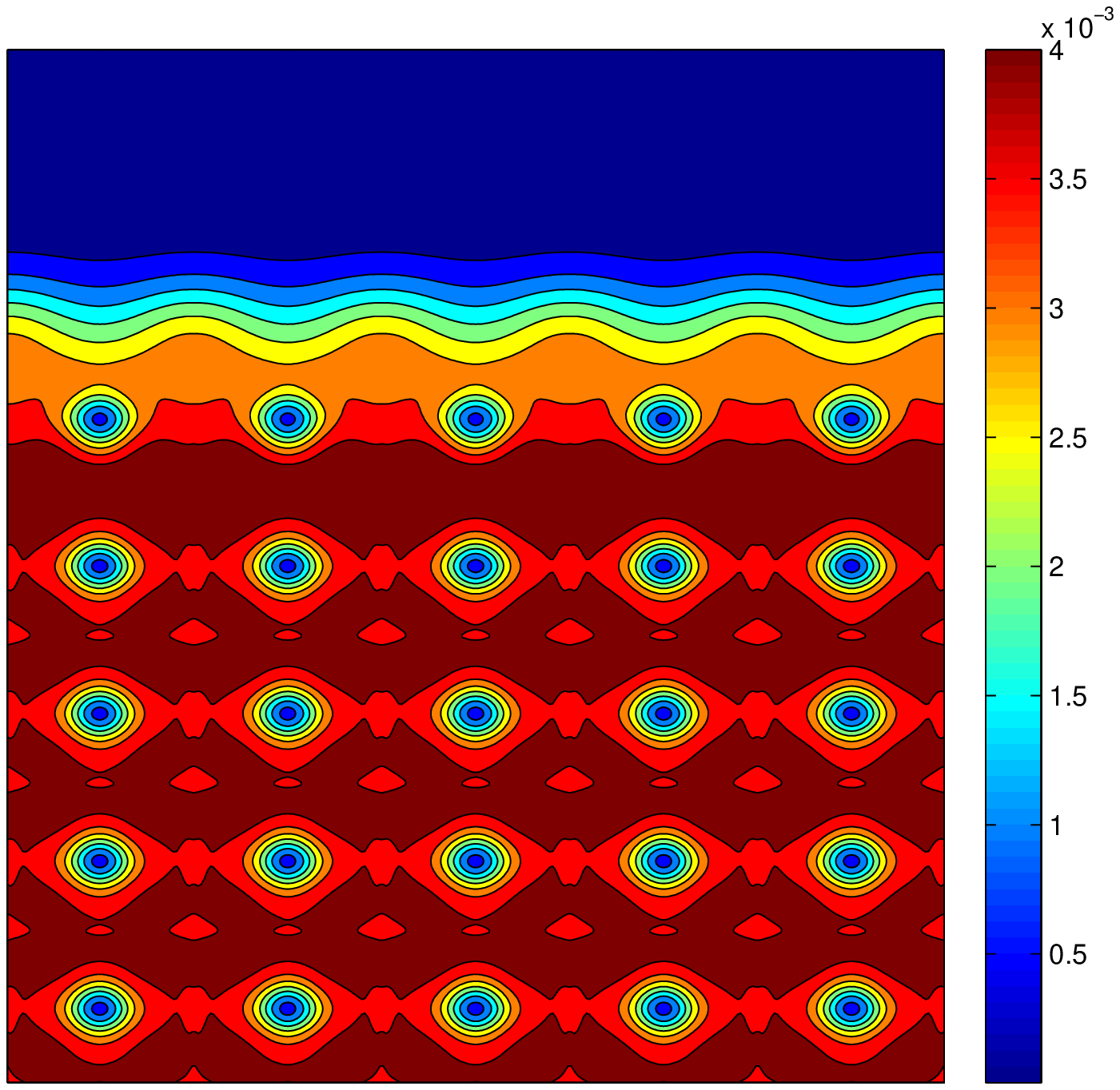}}
\subfloat[Edge plane]
{\label{Fig:ElectronDensity:Surface:Face}
\includegraphics[keepaspectratio=true,width=0.48\textwidth]
{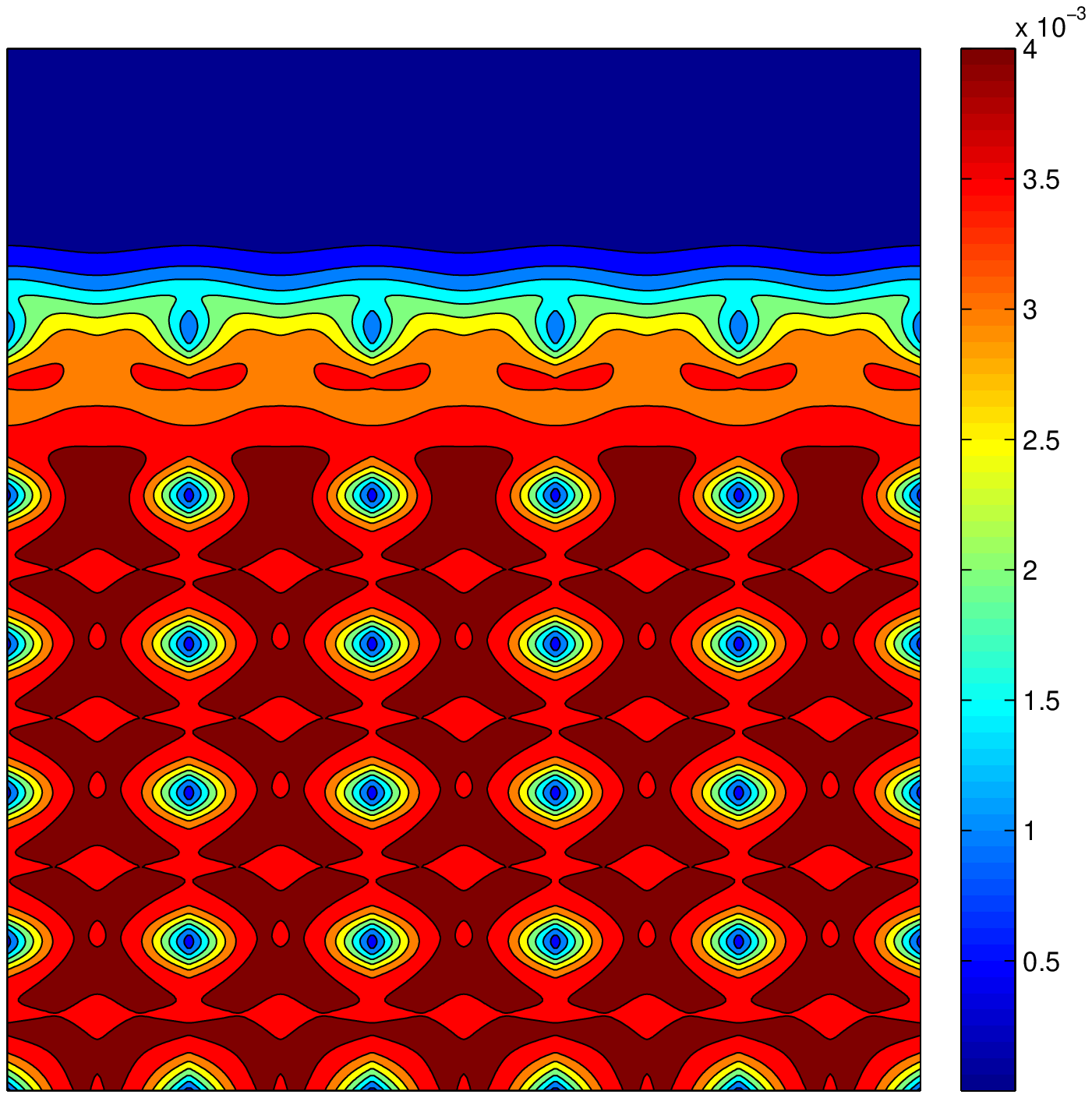}}
\caption{Electron density contours on slices through a  (001) surface of BCC sodium}
\label{Fig:ElectronDensity:Surface}
\end{figure}

We begin by verifying the convergence of the calculations with respect to the number of spectral quadrature points (K) in Fig.~\ref{Fig:Convergence_surface_Na}. The rapid convergence, highlighting the efficacy of the method, is noteworthy from the figure. The calculated surface energy and the displacement of the atoms are collected in Table \ref{table:LSSGQ:surface:sodium}. The calculated surface energy is in good agreement with previous values in literature \citep{Vitos1998}. Note that the forces on the atoms in the third layer and beyond are below the threshold value used for the calculations. Also, the convergence of the calculated quantities with respect to size of $\Omega_{RV}$ has been verified. Finally, the electron density contours on the mid plane and edge plane are plotted in Fig.~\ref{Fig:ElectronDensity:Surface} for purposes of illustration.

\begin{table}[h!]
\begin{center}
\caption{Properties of the (001) surface of BCC sodium}
  \label{table:LSSGQ:surface:sodium}
\begin{tabular}{lc}
    \hline
Surface energy  & 0.2 $J/m^2$ \\
Displacement of atoms  & \\
\hspace{3mm} First layer & 0.75 a.~u. \\
\hspace{3mm} Second layer & -0.19 a.~u.\\
\hline
\end{tabular}
\end{center}
\end{table}

\section{Coarse-Graining Density Functional Theory} \label{Sec:CGDFT}

{  Defects in crystalline solids involve elastic and electrostatic fields that may decay extremely slowly.   Therefore, accurate computations of such problems require very large computational domains, and even linear-scaling methods can be insufficient to fully understand them.  However, the slow decay offers an opportunity to approximate the problem in a manner that reduces the computational complexity with limited loss of accuracy.  In this section, we develop a method wherein we solve defects (using DFT) without the introduction of additional equations or uncontrolled approximations, but with substantially reduced effort.  This method is in the spirit of the quasicontinuum approximation used in atomistic models (\cite{Tadmor1996}) and OFDFT (\cite{Gavini:2007p340}), but also has important differences.  For definiteness, we discuss the method in terms of the finite-difference approximation scheme. We formulate the coarse-grained DFT (CGDFT) approach in Section \ref{Sec:CGKSDFT:Formulation} and validate it through examples in Section \ref{Sec:CGKSDFT:Validation}.}

\subsection{Formulation} \label{Sec:CGKSDFT:Formulation}
{ Consider the implementation of the LSSGQ method within the context of finite-differences (Section \ref{Sec:LSSGQ:FD}). Let us denote the collection of all finite-difference nodes $\{ \bx^p \}_{p=1}^{N_d}$ by $\mathcal{N}$. In the finite-difference approximation, $N_e$, $U_{band}$ and $S$ can be represented as the sums of nodal values (cf. Eqns. \ref{Eqn:FD:FermiEnergy}, \ref{Eqn:FD:BandEnergy}, \ref{Eqn:FD:Entropy}):
\begin{equation} \label{Eqn:sums}
N_e = h^D \sum_{p=1}^{N_d} \rho^{p} , \quad
U_{band} = \sum_{p=1}^{N_d} u^p, \quad S = \sum_{p=1}^{N_d} s^p \,,
\end{equation}
where $u$ and $s$ are used to denote the pointwise band structure energy and pointwise entropy respectively. Further, the nodal values $\rho^p$, $u^p$ and $s^p$ depend only on values of the spectral quadrature points $\{ \lambda^p_k \}_{k=1}^K$ and weights $\{ w_k^p \}_{k=1}^{K}$ associated with the $p^{th}$ node. (cf. Eqns. \ref{Eqn:FD:ElectronDensity}, \ref{Eqn:up}, \ref{Eqn:sp})
\begin{eqnarray}
\rho^{p} &=& \frac{2}{h^D} \sum_{k=1}^{K} w_k^p g(\lambda_k^p,\lambda_f), \label{Eqn:Nodal:rho} \\
u^p & = &  2 \sum_{k=1}^{K} w_k^p \lambda_k^{p} g(\lambda_k^p,\lambda_f), \label{Eqn:Nodal:Band} \\
s^p & = & -2 k_B \sum_{k=1}^{K} w_{k}^{p} [g(\lambda_k^{p},\lambda_f) \log g(\lambda_k^{p},\lambda_f) + (1-g(\lambda_k^{p},\lambda_f)) \log (1-g(\lambda_k^{p},\lambda_f))]. \label{Eqn:Nodal:Entropy}
\end{eqnarray}
Finally, recall that we can compute $\{\lambda^p_k\}_{k=1}^K$ and $\{w^p_k\}_{k=1}^K$ associated with a particular node independently of all the other nodes.  This observation provides the key idea behind coarse-graining.  

When a defect is introduced in an otherwise perfect lattice, we expect complex fluctuations in the electron density and other fields in the vicinity of the defect, but relatively smooth decay away from the defect.  Specifically, we expect the atomic displacements to decay in some polynomial fashion and the perturbation in the electron density, pointwise band structure energy and pointwise entropy to decay to zero away from the defect.  We wish to take advantage of this long range decay structure to develop an accurate reduced representation. Below, we discuss in detail how this is achieved. }

\subsubsection{Atomic positions}

We select representative atoms and interpolate the displacement for the remaining atoms as is done in the quasicontinuum method (cf., e.~g., \cite{Tadmor1996, Knap2001}). The density of the representative atoms is chosen so as to be high in the vicinity of the defect and to decrease with increasing distance to the defect. This varying resolution can be achieved through a variety of adaptive refinement schemes \citep{Tadmor1996, Knap2001}. 

{
\subsubsection{Electron density, band structure energy and entropy}
Since we expect the perturbation (due to defect) in the electron density, pointwise band structure energy and pointwise entropy to decay to zero, we seek an approximation scheme with the following decomposition
\begin{equation} \label{0andd}
\rho = \rho_0 + \rho_d, \quad u = u_0 + u_d, \quad s = s_0 + s_d
\end{equation}
where $\rho_0, u_0$ and $s_0$ are piecewise periodic, while $\rho_d, u_d$ and $s_d$ decay in a smooth fashion away from the defect. $\rho_0, u_0$ and $s_0$ can be determined by an inexpensive periodic calculation. Since $\rho_d, u_d$ and $s_d$ decay smoothly, we can develop an adaptive method for computing them. To this end, we pick representative nodes, the collection of which we denote by $\mathcal{R}$. The density of finite-difference nodes in $\mathcal{R}$ is chosen to be high in the vicinity of the defect and to progressively decrease with increasing distance from the defect. We then use the LSSGQ method to evaluate the spectral Gauss quadrature points and weights only at the representative nodes $\{ \bx^p, p \in \mathcal{R} \}$. We then evaluate the Fermi energy by solving for the constraint (Eqn. \ref{Eqn:sums}) with the interpolation scheme 
\begin{equation} \label{Eqn:rhoapprox}
\rho^{p} = \rho_{0}^p + \sum_{q \in \mathcal{R}} \gamma_{q}^{p} (\rho^q - \rho_{0}^q) .
\end{equation}
for the nodes $\{ \bx^p, p \in \mathcal{N}-\mathcal{R} \}$. $\gamma_{q}^{p}$ are weights decided by the degree of interpolation/approximation used. For the examples in the next section, we use cubic spline interpolations to determine the weights (\cite{Knott2000}). Once we locate the Fermi energy, we calculate the electron density at all the finite-difference nodes using Eqn. \ref{Eqn:Nodal:rho} for the nodes $\{ \bx^p, p \in \mathcal{R} \}$ and Eqn. \ref{Eqn:rhoapprox} for the nodes $\{ \bx^p, p \in \mathcal{N}-\mathcal{R} \}$. The free energy can be determined by evaluating the nodal band structure energy, nodal entropy using Eqns. \ref{Eqn:Nodal:Band}, \ref{Eqn:Nodal:Entropy} and using the interpolation
\begin{eqnarray} 
u^p & = & u_{0}^p + \sum_{q \in \mathcal{R}} \gamma_{q}^{p} (u^q - u_{0}^q) \label{Eqn:uapprox}\\
s^p & = & s_{0}^p + \sum_{q \in \mathcal{R}} \gamma_{q}^{p} (s^q-s_{0}^q) \label{Eqn:sapprox}.
\end{eqnarray}
for nodes $\{ \bx^p, p \in \mathcal{N}-\mathcal{R} \}$. Since we expect $\rho_d$, $u_d$ and $s_d$ to smoothly decay to zero away from the defect, we can use fewer and fewer representative finite-difference nodes away from the defect to capture these perturbations. As a consequence, the method will display sub-linear scaling with respect to the number of atoms. Therefore, this method opens the possibility of studying systems of much larger size compared to the linear-scaling LSSGQ method. 
}

\subsubsection{Electrostatic potential}
We write the electrostatic potential 
\begin{equation}
\phi = \phi_{0} + \phi_d
\end{equation}
and evaluate the perturbation $\phi_d$ by solving the equation
\begin{equation}  \label{cges}
-\frac{1}{4\pi} \nabla^2 \phi_d = \rho - \rho_{0} + b - b_0.
\end{equation}
By its very nature, $\phi_d$ is localized near the defect and decays to zero away from it. Therefore, it can be easily calculated by using any adaptive discretization tailored to this feature.
\vspace{\baselineskip}

{
We summarize the CGDFT method in Algorithm \ref{Algo:CGDFT}.
\newline \newline
\begin{algorithm}[h!] 
Generate guess for positions of the representative atoms ($\bR$) \\
\Repeat(Relaxation of representative atoms){Energy minimized with respect to positions of representative atoms}
{
        Calculate charge density of the nuclei ($b$) \\
        Generate piecewise periodic $\rho_0$ and $\phi_0$ \\
        Generate guess for electron density ($\rho$) \\
	\Repeat(Self-Consistent loop: SCF ){Convergence of self-consistent iteration}
        {
                  Calculate electrostatic potential ($\phi$) by solving Eqn.~\ref{cges} \\
		              Calculate exchange correlation potential ($V_{xc}$) \\
		              Calculate spectral quadrature points and weights at representative finite-difference nodes\\
                  Calculate Fermi energy ($\lambda_{f}$) using Eqn.~\ref{Eqn:sums} (utilizing interpolation given by Eqn. \ref{Eqn:rhoapprox})\\
                  Calculate electron density at representative nodes ($\rho^p, p \in \mathcal{R}$) using Eqn.~\ref{Eqn:Nodal:rho} \\
                  Interpolate the electron density to nodes ($\rho^p, p\in \mathcal{N} - \mathcal{R}$) using Eqn.~\ref{Eqn:rhoapprox}\\
                  Update the electron density (mixing) \\
        } 
        Calculate the forces on the nuclei using Eqn.~\ref{Eqn:Force:Nuclei}\\
}
       Evaluate free energy $\mathcal{F}$  using Eqns. \ref{Eqn:sums}, \ref{Eqn:Nodal:rho}, \ref{Eqn:Nodal:Band}, \ref{Eqn:Nodal:Entropy}, \ref{Eqn:uapprox} and \ref{Eqn:sapprox} for the band structure energy and entropy.
\caption{CGDFT method}
\label{Algo:CGDFT}
\end{algorithm}
}

\subsection{Validation} \label{Sec:CGKSDFT:Validation}

In this section, we verify the CGDFT method by applying it to a one-dimensional model problem and a Kohn-Sham surface relaxation problem, in Sections \ref{Sec:QC:Validation:1d} and \ref{Sec:QC:Validation:SurfaceRelaxation} respectively.

\subsubsection{One-dimensional model} \label{Sec:QC:Validation:1d}

\begin{figure}[h!]
\centering
\subfloat{
\includegraphics[keepaspectratio=true,width=0.48\textwidth]
{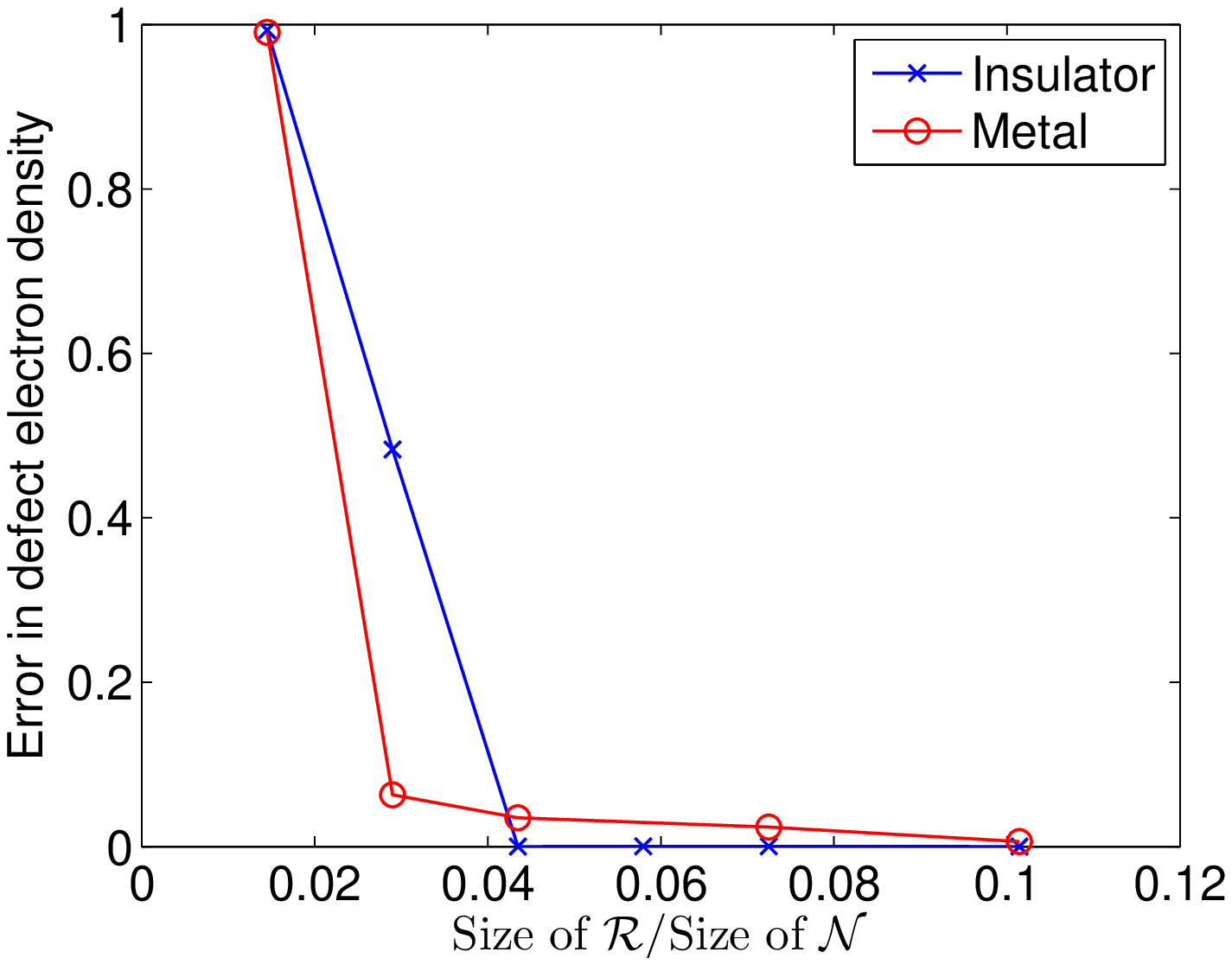}}
\subfloat {\includegraphics[keepaspectratio=true,width=0.48\textwidth]
{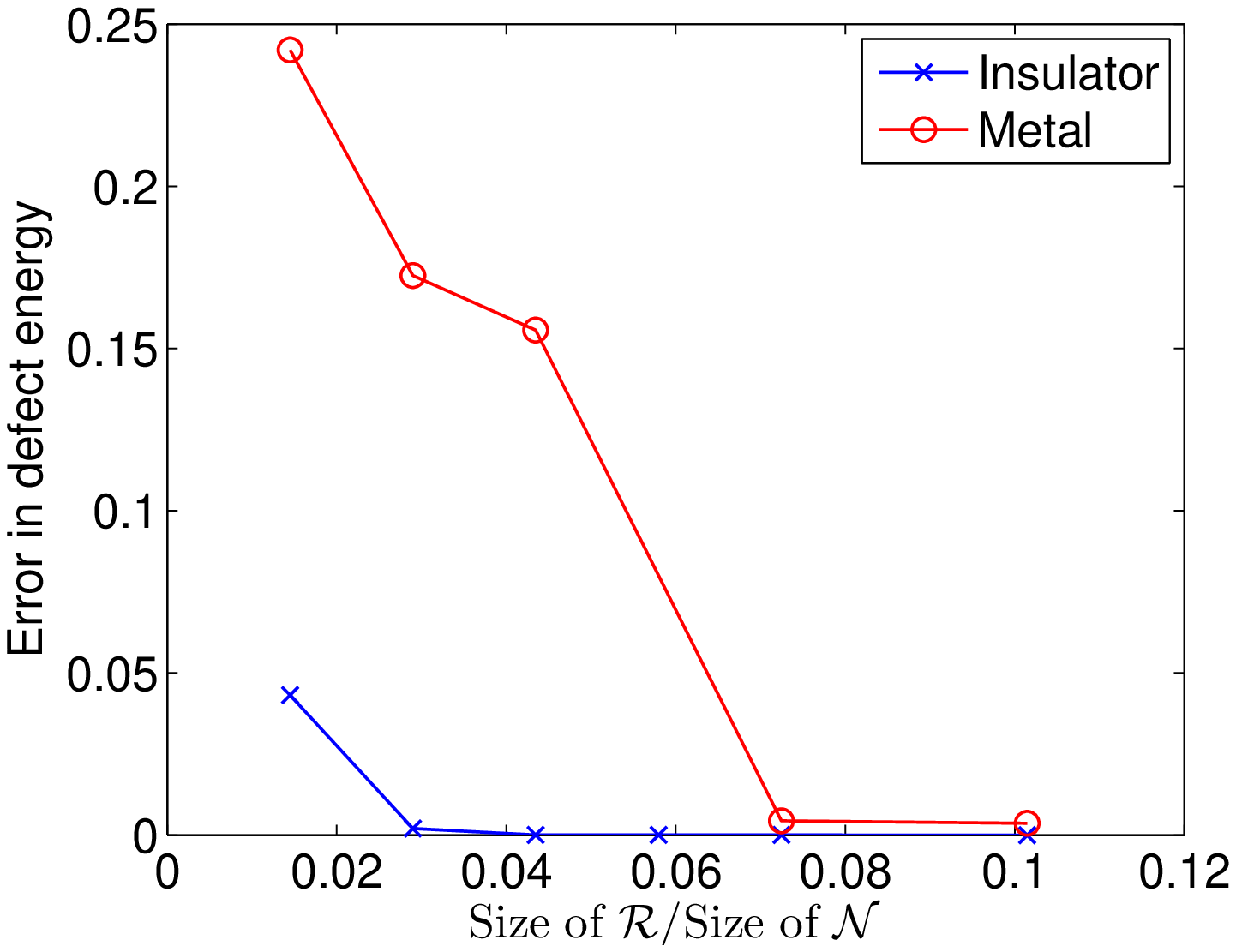}}
\caption{Convergence of the CGDFT method for the one-dimensional model problem with a point defect}
\label{Fig:QC:Convergence}
\end{figure}

We study the efficiency of CGDFT in the framework of the one-dimensional model presented in Section \ref{Sec:LSSGQ:Numerical:1d}. Consider a chain consisting of $101$ atoms with unit spacing. We introduce a point defect by removing the center atom. We coarse-grain in both directions from the position of the defect and adopt the procedure described in Section \ref{Sec:CGKSDFT:Formulation} to solve for the electron density and the energy. We plot the error in the defect energy and defect electron density so obtained, as a function of the number of representative nodes in Fig.~\ref{Fig:QC:Convergence}. The error is measured and normalized with respect to the fully resolved LSSGQ solution. It is clear that we get rapid convergence in the solution for both metals and insulators. As expected, convergence in the case of insulators is much more rapid owing to the highly localized nature of the defect perturbation. Therefore, CGDFT can be utilized to solve the defect problem at a small fraction of the original computational cost, without any significant loss of accuracy.

\subsubsection{Kohn-Sham problem: Surface relaxation} 
\label{Sec:QC:Validation:SurfaceRelaxation}

We now study the (001) surface relaxation of BCC sodium using CGDFT, previously analyzed using the LSSGQ method in Section \ref{Sec:LSSGQ:Validation:KS}. We employ the procedure described in Section \ref{Sec:LSSGQ:Validation:KS} with the introduction of coarse-graining described in Section \ref{Sec:CGKSDFT:Formulation}.  However, we do not introduce coarse-graining for the electrostatic potential since the computational time involved in its calculation is negligible. We compare the results obtained using CGDFT with the fully resolved solution obtained using the LSSGQ method. A summary of the results is presented in Table \ref{table:CGDFT}. We can see that there are significant computational savings when using CGDFT at no appreciable loss of accuracy. It should be carefully noted that the surface relaxation problem is effectively one-dimensional for purposes of coarse-graining. As a consequence, the computational savings accrued are likely to be less substantial than in fully two or three dimensional problems.

\begin{table}[h!]
\begin{center}
\caption{Summary of the results for CGDFT}
\label{table:CGDFT}
\begin{tabular}{ll}
\hline
Number of representative nodes/Total number of nodes &  0.25 \\
Number of representative atoms/Total number of atoms & 0.3 \\
Normalized error in surface energy & 0.12 \% \\
Normalized error in displacement of atoms & \\
\hspace{3mm} First layer & 0.11 \% \\
\hspace{3mm} Second layer & 0.17 \% \\
\hline
\end{tabular}
\end{center}
\end{table}

\section{Conclusions} \label{Sec:Conclusions}

We have developed a real-space formulation for coarse-graining DFT (CGDFT). Traditional implementations of DFT solve for the the orbitals, which results in cubic-scaling with respect to the number of atoms. Furthermore, they are not amenable to coarse-graining because of the global nature of the orthonormality constraint. In order to overcome this difficulty, we have developed a linear-scaling method for DFT (LSSGQ), where we directly evaluate the electron density without evaluating the individual orbitals. We accomplish this direct evaluation by performing Gauss quadrature over the spectrum of the linear Hamiltonian operator in each iteration of the SCF method. As a second approximation step, we have appended coarse-graining approximations to the LSSGQ method. Specifically, we decompose the solution into a periodic component and the perturbation caused by lattice defects. Since the perturbation is expected to decay rapidly away from the defect, fine resolution in only needed in the vicinity of the defect and the resolution can be coarse-grained elsewhere. This coarse-graining enables the analysis of defects at a fraction of the computational cost, without any appreciable loss of accuracy. We have validated the LSSGQ and CGDFT methods through a number of examples of application, the results of which are in good agreement with literature values. Furthermore, we have verified the convergence of the CGDFT solution to the fully resolved solution. These verification and validation examples, taken together, show CGDFT to be highly transferable, efficient and accurate.

We close by pointing to some outstanding issues and limitations of the CGDFT method and to possible extensions of the method thereof. A principal shortcoming of the implementation described in this paper is the non-automated nature of the coarse-graining approximations. An optimal and automated coarse-graining technique for CGDFT would greatly extend the range, scope and robustness of the method. In addition, a Rayleigh-Ritz or Galerkin spatial discretization scheme, e.~g., based on finite elements, would endow the approximation scheme with a variational structure, thereby opening the way to energy-based adaptive methods. In addition, the convergence of variational approximation schemes can be established by means of powerful tools such as $\Gamma$-convergence. However, it should be noted that finite-element bases lack orthogonality in general and result in generalized eigenvalue problems. This property may potentially add to the computational cost and implementational complexity associated with the method. These issues are currently being investigated by the authors. 

\section*{Acknowledgements}
This work draws from the thesis of PS at the California Institute of Technology.  We gratefully acknowledge the support of the 
US Army Research Office (under MURI grant number W911NF-07-1-0410), the US Department of Energy National Nuclear Security Administration (under Award Number DE-FC52-08NA28613 through Caltech's ASC/PSAAP Center for the Predictive Modeling and Simulation of High Energy Density Dynamic Response of Materials) and the US National Science Foundation (under PIRE grant number OISE-0967140).

\begin{appendix}

\section{Pad{\'e} approximation and recursion method} \label{Sec:RM}

The recursion method \citep{Haydock1980} has found many applications in the physics literature. Here, we discuss the close connection between the LSSGQ and Recursion methods. In doing so, we also establish their relationship to the Pad{\'e} approximation. Details of the notation not described here can be found in Sections \ref{Sec:MathematicalBackground} and \ref{Sec:LSSGQ}.

Let $\mathcal{R_{H}}(z) = (z\mathcal{I} - \mathcal{H})^{-1}$ represent the resolvent of the operator $\mathcal{H}$ for $z \in \rho(\mathcal{H}) = \mathbb{C} \setminus \sigma(\mathcal{H})$ where $\rho(\mathcal{H})$ denotes the resolvent set of $\mathcal{H}$ and $\mathbb{C}$ is the set of all complex numbers. $\mathcal{R_{H}}(z)$ is holomorphic $\forall z \in \rho(\mathcal{H})$. For $z \in \rho(\mathcal{H})$, $\zeta \in H$ it follows
\begin{eqnarray}
\mathcal{R_{\mathcal{H}}}(z) & = & \int_{\sigma(\mathcal{H})} \frac{1}{z-\lambda} \, d\mathcal{E} (\lambda) \,, \\
G_{\zeta,\zeta}(z) = (\mathcal{R_{\mathcal{H}}}(z) \zeta, \zeta) & = & \int_{\sigma(\mathcal{H})} \frac{1}{z-\lambda} \, d\mathcal{E}_{\zeta,\zeta} (\lambda). \label{Eqn:SD:R}
\end{eqnarray}
Taking the series expansion of $G_{\zeta,\zeta}(z)$ around the point $z=\infty$, we obtain
\begin{eqnarray}
G_{\zeta,\zeta} (z) & = & \sum_{k=0}^{\infty} \frac{1}{z^{k+1}} \left( \mathcal{H}^{k} \zeta, \zeta \right) \nonumber \\
                  & = & \sum_{k=0}^{\infty} \frac{1}{z^{k+1}} \int_a^b \lambda^{k} \, d \mathcal{E}_{\zeta,\zeta} (\lambda) = \sum_{k=0}^{\infty} \frac{\mu_{k}}{z^{k+1}} \label{Eqn:series:moments}
\end{eqnarray}
which converges for $|z| > \norm{\mathcal{H}}$.

Diagonal Pad\'{e} approximants for $G_{\zeta,\zeta}(z)$ are rational functions of the form $q_K(z)/p_K(z)$ which satisfy the following property
\begin{equation} \label{Eqn:Pade:approx:series}
\left( G_{\zeta,\zeta}(z) - \frac{q_K(z)}{p_K(z)} \right) = \mathcal{O} \left( \frac{1}{z^{2K+1}} \right)
\end{equation}
and are locally the best rational approximations for a given power series like Eqn.~\ref{Eqn:series:moments} (cf., e.~g., \cite{Suertin2002}). The denominator polynomials $p_K(z)$, are a set of orthogonal polynomials with respect to the measure $\mathcal{E}_{\zeta,\zeta}$ (cf., e.~g., \cite{Moren2008})
\begin{equation} \label{Eqn:Pade:Den:Orthogonal}
\int_a^b \lambda^{k} p_K(\lambda) \, d\mathcal{E}_{\zeta,\zeta}(\lambda) = 0\,, \quad k=0,1, \ldots, K-1
\end{equation}
and the numerator polynomial $q_K(z)$ can be expressed in terms of $p_K(z)$ as follows
\begin{equation} \label{Eqn:Pade:Den:Num:Relation}
q_K(z) = \int_a^b \frac{p_K(z)-p_K(\lambda)}{z-\lambda} \, d\mathcal{E}_{\zeta,\zeta}(\lambda).
\end{equation}
The Pad\'{e} approximants are constructed using the coefficients of the power series and provide an efficient analytic continuation of the series beyond the domain of convergence, which in the present case is given by $|z| > \norm{\mathcal{H}}$. The convergence of diagonal Pad\'{e} approximants follows from Markov's theorem (cf., e.~g., \cite{Suertin2002}), whereby any Markov function like $G_{\zeta,\zeta}(z)$ can be recovered in $\rho(\mathcal{H})$ from the coefficients of the Laurent expansion of the function at the point $z=\infty$ (i.e. from the moments of the measure $\mathcal{E}_{\zeta,\zeta}$).

The orthogonal polynomials $\{p_k\}_{k=1}^{K}$ can be generated via the recurrence relation (cf., e.~g., \cite{Golub2010})
\begin{eqnarray} \label{Eqn:RecurrenceRelation:orthogonal}
p_{k+1}(\lambda) = (\lambda - a_{k+1})p_k(\lambda) - b_k^{2} p_{k-1}(\lambda) \,, \quad k=0, 1, \ldots, K-1 \nonumber \\
p_{-1}(\lambda) = 0 \,, \quad p_0(\lambda) = 1
\end{eqnarray}
where
\begin{eqnarray}
a_{k+1} & = & \frac{\langle \lambda p_k, p_k \rangle_{\zeta}}{\langle p_k, p_k \rangle_{\zeta}} \,, \quad k = 0, 1, \ldots, K-1 \nonumber \\
b_k^{2} & = & \frac{\langle p_k, p_k \rangle_{\zeta}}{\langle p_{k-1}, p_{k-1} \rangle_{\zeta}} \,, \quad k = 1, 2, \ldots, K-1.
\end{eqnarray}
Using Eqns. \ref{Eqn:Pade:Den:Orthogonal} and \ref{Eqn:Pade:Den:Num:Relation}, we see that $\{q_k\}_{k=1}^{K}$ can be generated with a similar recurrence relation as that for $\{p_k\}_{k=1}^{K}$ but with different initializing conditions
\begin{eqnarray}
q_{k+1}(\lambda) = (\lambda - a_{k+1})q_k(\lambda) - b_k^{2} q_{k-1}(\lambda) \,, \quad k = 0, 1 , \ldots, K-1 \nonumber \\
q_{-1}(\lambda) = -1, \,\, q_{0}(\lambda) = 0 \,\, \mathrm{and} \,\, b_0^{2} = 1 .
\end{eqnarray}

Corresponding to these recurrence relations we have the tridiagonal matrix
\begin{eqnarray}
J_K =  \left( \begin{array}{ccccc}
a_1 & 1 & & & \\
b_1^{2} & a_2 & 1 & & \\
 & \ddots & \ddots & \ddots & \\
 & & b_{K-2}^{2} & a_{K-1} & 1 \\
 & & & b_{K-1}^{2} & a_K
\end{array} \right)
\end{eqnarray}
Expanding $\mathrm{det} (zI_k - J_k)$ with respect to its last row, we obtain the recurrence formula for the determinant
\begin{eqnarray}
\mathrm{det} (zI_{k+1}-J_{k+1}) = (z-a_{k+1})\mathrm{det}(zI_k-J_k) - b_k^{2} \mathrm{det} (zI_{k-1}-J_{k-1}) \,, k=0, 1, \ldots, K-1 \nonumber \\
\mathrm{det}(zI_{-1}-J_{-1}) = 0 \,, \quad \mathrm{det}(zI_0-J_0) = 1
\end{eqnarray}
which is the same as that satisfied by $\{p_k\}_{k=0}^{K-1}$ as well as $\{ \mathrm{det}(zI_k-\hat{J}_k) \}_{k=0}^{K-1}$. Consequently
\begin{equation}
p_K(z) = \mathrm{det} (z I_K - J_K) = \mathrm{det} (z I_K - \hat{J}_K).
\end{equation}
Similarly
\begin{equation}
q_K(z) = \mathrm{det} (z I_{K-1} - J_{K-1}^{1}) = \mathrm{det} (z I_{K-1} - \hat{J}_{K-1}^{1})
\end{equation}
where the superscript $i$ is used to denote the matrix obtained by deleting the first $i$ rows and columns. Therefore
\begin{equation} \label{Eqn:PA:det:ratio}
\frac{q_K(z)}{p_K(z)} = \frac{\mathrm{det} (z I_{K-1} - \hat{J}_{K-1}^{1})}{\mathrm{det} (z I_K - \hat{J}_K)} = e_1(zI_K-\hat{J}_K)^{-1}e_1.
\end{equation}
Above, the first column of $I_K$ is denoted by $e_1$.

Expanding $\mathrm{det}(zI_K-\hat{J}_K)$ about the first row or column we obtain from Eqn.~\ref{Eqn:PA:det:ratio}
\begin{eqnarray}
\frac{q_K(z)}{p_K(z)} = \cfrac{1}{z-a_1-\cfrac{\mathrm{det}(zI_{K-2}-\hat{J}_{K-2}^{2})}{\mathrm{det}(zI_{K-2}-\hat{J}_{K-2}^{1})}}
\end{eqnarray}
and continuing similarly we obtained a continued fraction representation for the Pad{\'e} approximant
\begin{equation}
\frac{q_K(z)}{p_K(z)} = \cfrac{1}{z-a_1-\cfrac{b_1^2}{z-a_2-\ldots - \cfrac{b_{K-1}^2}{z-a_K}}}
\end{equation}
In the recursion method, the above expression is used to find the projected density of states \citep{Haydock1980} from which all the necessary quantities are evaluated. Since the projected density of states has poles at the zeros of $p_K(z)$, a number of techniques to smoothen it have been developed. These include either using a terminator, simplest one being $\{a_k\}_{k=K+1}^{\infty} = a_{\infty}$, $\{b_k\}_{k=K}^{\infty} = b_{\infty}$ or evaluating the Pad{\'e} approximant slightly away from the real axis. However, these are uncontrolled approximations and can sometimes lead to inaccuracies.

Finally, we look at the connection between Pad{\'e} approximation and Gauss quadrature (cf., e.~g., \cite{Assche2006}). Multiplying both sides of Eqn.~\ref{Eqn:Pade:approx:series} by a polynomial of degree at most $2K-1$ denoted by $\pi_{2K-1}(z)$, integrating along a contour $\Gamma$ encircling the real line and subsequently using Eqn.~\ref{Eqn:SD:R} we obtain
\begin{equation}
\int_a^b \pi_{2K-1}(\lambda) \, d\mathcal{E}_{\eta,\eta}(\lambda) = \sum_{j=1}^{K} w_j \pi_{2K-1}(\lambda_j)
\end{equation}
where
\begin{equation}
w_j = \frac{q_K(\lambda_j)}{p_K^{'}(\lambda_j)}
\end{equation}
is the residue of the Pad{\'e} approximant at the zeros $\lambda_j$ of $p_K$. It is clear that all polynomials of degree $2K-1$ are integrated exactly, therefore establishing the connection between the Pad{\'e} approximation, recursion method and the LSSGQ method.

\section{LSSGQ method applied to systems with periodicity} \label{App:LSSGQ:periodic}

In this section, we discuss the application of the LSSGQ method to the study of periodic systems. Typically, this would involve the need for Bloch periodic boundary conditions on the orbital. However, since we directly evaluate the electron density and not the individual orbitals, we can circumvent the need for Bloch periodic boundary conditions using the procedure described below. For clarity, we discuss the procedure in terms of the finite-difference approximation. However, the method is not restricted by the choice of basis.

First, we specify a representative volume $\Omega_{RV}$ which has the property that the solution for the entire system can be ascertained through a translational mapping of the solution obtained in $\Omega_{RV}$. We use a finite-difference grid of uniform spacing $h$ to discretize $\Omega_{RV}$, which we denote by $\chi_{RV}$. Next, we define an extended volume $\Omega_{EV} \supset \Omega_{RV}$, which is discretized using a finite-difference grid of spacing $h$, denoted by $\chi_{EV} \supset \chi_{RV}$. We evaluate the spectral quadrature nodes and weights for $\chi_{RV}$ using the procedure outlined in Section \ref{Sec:LSSGQ:FD}. It is important to note that the size of $\Omega_{EV}$ is chosen such that all the vectors in the Lanczos iteration given by Eqn.~\ref{Eqn:LanczosIteration:FD} have nonzero components only for finite-difference nodes inside $\Omega_{EV}$. Since the initial vector $\bf{v}_0$ always has a single nonzero entry, the required size of $\Omega_{EV}$ can be easily calculated using the bandwidth of the discretized $\mathcal{H}$ and the number of spectral quadrature points used.

\end{appendix}

\bibliographystyle{elsarticle-harv}
\bibliography{R-KS-DFT_jmps}

\end{document}